\documentclass[]{spie}  

\usepackage{amsmath,amsfonts,amssymb}
\usepackage{graphicx}
\usepackage[colorlinks=true, allcolors=blue]{hyperref}

\usepackage{wrapfig}
\usepackage{sidecap}

\title{Polarization Modeling and Predictions for DKIST Part 2: Application of the Berreman Calculus to Spectral Polarization Fringes of Beamsplitters and Crystal Retarders.}

\author[a]{David M. Harrington}
\author[b]{Frans Snik}
\author[b]{Christoph U. Keller}
\author[c]{Stacey R. Sueoka}
\author[d]{Gerard van Harten}
\affil[a]{National Solar Observatory, 8 Kiopa'a Street, Ste 201 Pukalani, HI 96768, USA}
\affil[b]{Leiden Observatory, Leiden University, P.O. Box 9513, 2300 RA Leiden, The Netherlands}
\affil[c]{National Solar Observatory, 3665 Discovery Drive, Boulder, CO, 80303, USA}
\affil[d]{Jet Propulsion Laboratory, M/S 233-200, 4800 Oak Grove Drive, Pasadena, CA 91109}

\authorinfo{Further author information: (Send correspondence to D.M.H) \\
D.M.H.: E-mail: dharrington@nso.edu, Telephone: 1 808 572 6888}

\begin{document} 
\maketitle

\begin{abstract}

We outline polarization fringe predictions derived from a new application of the Berreman calculus for the Daniel K. Inouye Solar Telescope (DKIST) retarder optics. The DKIST retarder baseline design used 6 crystals, single-layer anti-reflection coatings, thick cover windows and oil between all optical interfaces. This new tool estimates polarization fringes and optic Mueller matrices as functions of all optical design choices. The amplitude and period of polarized fringes under design changes, manufacturing errors, tolerances and several physical factors can now be estimated. This tool compares well with observations of fringes for data collected with the SPINOR spectropolarimeter at the Dunn Solar Telescope using bi-crystalline achromatic retarders as well as laboratory tests. With this new tool, we show impacts of design decisions on polarization fringes as impacted by anti-reflection coatings, oil refractive indices, cover window presence and part thicknesses. This tool helped DKIST decide to remove retarder cover windows and also recommends reconsideration of coating strategies for DKIST. We anticipate this tool to be essential in designing future retarders for mitigation of polarization and intensity fringe errors in other high spectral resolution astronomical systems. 

\end{abstract}


\section{DKIST OPTICS}
\label{sec:intro}  

The Daniel K. Inouye Solar Telescope (DKIST) on Haleakal\={a}, Maui, Hawai'i is presently under construction with operations beginning around 2020.  The telescope has a 4.2m off-axis f/2 primary mirror (4.0m illuminated) and a suite of polarimetric instrumentation in a coud\'{e} laboratory \cite{2014SPIE.9145E..25M, Keil:2011wj, Rimmele:2004ew}.  Many of the proposed science cases rely on high spectral resolution polarimetry.  In many astronomical spectropolarimeters, fringes in intensity and polarization are the dominant source of error. With this paper, we adapt a formalism common in the thin-film industry and apply it to many-crystal retarders. We show how to predict fringe properties as well as to anticipate their response during the instrument design process. 

DKIST uses seven mirrors to feed light to the coud\'{e} lab \cite{Marino:2016ks, McMullin:2016hm,Johnson:2016he,2014SPIE.9147E..0FE, 2014SPIE.9147E..07E, 2014SPIE.9145E..25M}. Operations involve four polarimetric instruments presently spanning the 380nm to 5000nm wavelength range. At present design, three different retarders are in fabrication for use in calibration near the Gregorian focus \cite{2014SPIE.9147E..0FE,Sueoka:2014cm,Sueoka:2016vo}.  A train of dichroic beam splitters allows for rapid changing of instrument configurations and simultaneous operation of three polarimetric instruments covering 380nm to 1800nm using the adaptive optics system \cite{2014SPIE.9147E..0FE, 2014SPIE.9147E..07E, 2014SPIE.9147E..0ES, SocasNavarro:2005bq}. Another instrument (CryoNIRSP) can receive all wavelengths to 5000 nm but without use of the adaptive optics system.

Complex polarization modulation and calibration strategies are required for such a mulit-instrument system \cite{2014SPIE.9147E..0FE,2014SPIE.9147E..07E, Sueoka:2014cm, 2015SPIE.9369E..0NS, deWijn:2012dd, 2010SPIE.7735E..4AD}.  The planned 4m European Solar Telescope (EST), though on-axis, will also require similar calibration considerations \cite{SanchezCapuchino:2010gy, Bettonvil:2011wj,Bettonvil:2010cj,Collados:2010bh}.  Many solar and night-time telescopes have performed polarization calibration of complex optical pathways \cite{DeJuanOvelar:2014bq, Joos:2008dg, Keller:2009vj,Keller:2010ig, Keller:2003bo, Rodenhuis:2012du, Roelfsema:2010ca, 1994A&A...292..713S, 1992A&A...260..543S,  1991SoPh..134....1A, Schmidt:2003tz, Snik:2012jw,  Snik:2008fh, Snik:2006iw, SocasNavarro:2011gn, SocasNavarro:2005jl, SocasNavarro:2005gv, Spano:2004ge, Strassmeier:2008ho, Strassmeier:2003gt, Tinbergen:2007fd, 2005A&A...443.1047B, 2005A&A...437.1159B}.  We refer the reader to recent papers outlining the various capabilities of the first-light instruments \cite{McMullin:2016hm, 2014SPIE.9147E..07E, 2014SPIE.9145E..25M, 2014SPIE.9147E..0FE, Rimmele:2004ew}. 

Berreman (1972) \cite{1972JOSA...62..502B} has formulated a 4x4-matrix method that describes electromagnetic wave propagation in biaxial media. The interference of forward and backward propagating electromagnetic waves inside birefringent media is included in this very general theory. This Berreman calculus can also be used to describe wave interference in multiple birefringent layers, crystals, chiral coatings and other complex optical configurations with many birefringent layers of arbitrary optical axis orientation. A recent textbook by McCall, Hodgkinson and Wu has further developed and applied the Berreman calculus \cite{McCall:2015uu}.  

In this paper, we call this reference MHW extensively. The Berreman formalism is very common in the thin-film coating and liquid-crystal industries. However, we show the formalism applies equally well to thick crystal retarders, beamsplitters, dichroics and other optics common in astronomical telescopes. We adopt this formalism and show the benefits of it's application to high spectral resolution astronomical instrument design and calibration. 

The MHW textbook comes with a MATLAB software demonstration kit that allows for exploration of the calculus. With this book, we developed our own software package (in Python) to apply this framework to crystal-based retarder optics, dichroic beam splitters and windows in use for DKIST instruments at resolving power of $\lambda$/$\delta\lambda$ of $>$200,000.   MHW \cite{McCall:2015uu} show how to calculate important optical quantities such as transmittance, polarizance and retardance in a wide range of coatings and crystals by using their 4x4-matrix method.  

In this paper, we apply Berreman's 4x4-matrix formulation to thick many-crystal wave plates to describe the polarized spectral fringes. We show how fringes that originate from these non-absorbing, uniaxial media and how we can use this new framework to make informed design decisions for anti-reflection coatings, refractive index matching oil layering, crystal thickness choices and other thermal issues associated with typical solar telescope heat loads on calibration optics.  Fringes change not only the transmission of the optic, but also change the retardance at the design wavelength and add all aspects of polarization to the optic  (transmission, diattenuation, polarizance, circular and linear retardance). 

A number of references dealing with aspects of polarized spectral fringes can be found in the astronomical literature. Semel (2003) discussed fringes in an intuitive way by summing (reflected) waves inside a wave plate \cite{2003A&A...401....1S}. Clarke has used the theory of a single Fabry-P\'erot interferometer to describe interference effects in single wave plates, and Heavens matrices \cite{Heavens:1965uq} to describe interference effects in compound, achromatic and Pancharatnam wave plates \cite{2004JOptA...6.1041C,2005A&A...434..377C,Clarke:2004gm,2004JOptA...6.1047C, Clarke:2009ty}.  Often, simple Fourier or function fitting techniques are used to remove the fringes with varying degrees of difficulty and success \cite{Aitken:2001ih, Harries:1996vf, Harrington:2015cq}

Although these theories describe aspects of the interference patterns, they are mostly ad-hoc theories with several limiting assumptions. The Berreman calculus contains all polarization phenomena and is much more general \cite{McCall:2015uu}. For example, the astronomical research articles consider only normal incidence and they become complicated when calculating the interference effects for multiple birefringent layers or thick crystals. The ad-hoc formalisms cannot accept arbitrarily oriented bi-axial materials. The Berreman calculus can compute all polarization phenomena for an arbitrary stack of crystals with an arbitrary number of coatings on those crystals with each crystal at any desired orientation and any desired incidence angle.  

The main limitation of the formalism is in the assumption of complete beam overlap between reflected and transmitted beams. In the Berreman formalism for a finite sized beam at non-normal incidence, the multiple reflections inside a thick plate will, in practice not overlap with the incoming beam. An additional limitation is the formalism does not include converging or diverging beams.  Modeling a real optical system requires additional consideration. Berreman always assumes infinite coherence lengths, and that all multiple reflections stay within the optical path. For most astronomical applications, this beam overlap assumption is reasonably valid and other scaling relations can be used to assess impact of converging beams.

Other instrument design strategies include designing retarders such that fringes are unresolved and average below detection limits \cite{Snik:2012jw, Snik:2015hs}.  Additional optics can be added to minimize fringe amplitudes and allow the use of wedged optics for fringe reduction \cite{2005A&A...437.1159B, Schmidt:2003tz}. We show here that the Berreman formalism is very powerful for modeling the DKIST retarder optics which consist of 6 birefringent crystals, 2 cover windows, 14 anti-reflection coatings and thin layers of index-matching oil.  Modeling the full polarization performance of such a retarder at high spectral resolution in the presence of manufacturing imperfections would be impossible without the Berreman formalsim. 

\begin{wrapfigure}{r}{0.40\textwidth}
\begin{equation}
{\bf M}_{ij} =
 \left ( \begin{array}{rrrr}
 II   	& QI		& UI		& VI		\\
 IQ 	& QQ	& UQ	& VQ	\\
 IU 	& QU	& UU	& VU		\\
 IV 	& QV	& UV		& VV		\\ 
 \end{array} \right ) 
\label{eqn:MM}
\end{equation}
\vspace{-4mm}
\end{wrapfigure}

In this work, we follow standard notation for propagation of polarization through an optical system.  The Stokes vector is denoted as {\bf S} = $[I,Q,U,V]^T$. The Mueller matrix is the 4x4 matrix that transfers Stokes vectors \cite{1992plfa.book.....C, Chipman:2014ta, Chipman:2010tn}. Each element of the Mueller matrix is denoted as the transfer coefficient \cite{Chipman:2010tn, 2013pss2.book..175S}. For instance the coefficient [0,1] in the first row transfers $Q$ to $I$ and is denoted $QI$. The first row terms are denoted $II$, $QI$, $UI$, $VI$. The first column of the Mueller matrix elements are $II$, $IQ$, $IU$, $IV$. In this paper we will use the notation in Equation \ref{eqn:MM}

We also will adopt a standard astronomical convention for displaying Mueller matrices.  We normalize every element by the $II$ element to remove the influence of transmission on the other matrix elements as seen in Equation \ref{eqn:MM_IntensNorm}.  Thus subsequent Figures will display a matrix that is not formally a Mueller matrix but is convenient for displaying the separate effects of transmission, retardance and diattenuation in simple forms.

\begin{wrapfigure}{l}{0.45\textwidth}
\vspace{-3mm}
\begin{equation}
 \left ( \begin{array}{rrrr}
 II   		& QI/II	& UI/II	& VI/II	\\
 IQ/II 	& QQ/II	& UQ/II	& VQ/II	\\
 IU/II 	& QU/II	& UU/II	& VU	/II	\\
 IV/II	 	& QV/II	& UV	/II	& VV	/II	\\ 
\end{array} \right ) 
\label{eqn:MM_IntensNorm}
\end{equation}
\vspace{-3mm}
\end{wrapfigure}

In this work we assume basic familiarity with the MHW textbook\cite{McCall:2015uu} and the basic thin film calculations by Abeles and Heavens matrices \cite{Heavens:1965uq}. 

We will first show fringe predictions for a simple isotropic anti-reflection coating and make comparisons with traditional optical modeling tools. We then show simple zero-order and multi-order crystal retarders compared to laboratory measurements.   We then show predictions and observations for a bi-crystalline achromat using the SPINOR spectropolarimeter at the Dunn Solar Telescope. We then end with prediction and analysis for the DKIST suite of six calibration retarders and polarization modulators. Each of these retarders were designed to be six crystals, anti-reflection coatings on all surfaces, index matching oil layers between all surfaces and cover windows.


\section{Single-layer isotropic MgF$_2$ coatings \& isotropic substrates}

In this section, we show how the Berreman calculus gives the same fringe results as other approaches in the limit of isotropic materials.  As an example, we compare the predicted reflectivity from a large commercial software package (Zemax) against our Berreman calculations.  We use a simple fused silica window and a single layer of isotropic amorphous magnesium flouride MgF$_2$ as an anti-reflection coating. 

To show some familiar effects using simple isotropic materials, we compute Mueller matrices, diattenuation and retardance for MgF$_2$ coatings of different thicknesses.  We show an uncoated fused silica window, a quarter-wave thick coating at a central wavelength of 600nm and a very thick coating of two waves thickness at 2000nm central wavelength. The thicker coating would physically be 2924nm thick and produce many polarization fringe cycles across visible wavelengths. The quarter-wave coating would have a 109nm physical thickness when assuming a refractive index of 1.370 at 600nm. 

We will describe here some of the basic effects using a language that only considers a single back-reflected beam from the coating-window interface. One could describe the effects more accurately as an infinite series of internal reflections, where considering only one reflection is an approximation that is reasonable only when the amplitudes are small.  As described in the Berreman calculus, a more technically accurate approach is to describe the problem as the electric field boundary conditions that satisfy propagation through an arbitrary stack of arbitrarily low or high reflectivity without restriction of isotropy or material orientation. The electric field must satisfy equations of continuity.  But for this simple case, we use the simple approximate language and give intuitive numbers that aim to highlight the differences in the approach. 

\begin{figure}[htbp]
\begin{center}
\vspace{-3mm}
\hbox{
\hspace{-2.0em}
\includegraphics[height=11.8cm, angle=0]{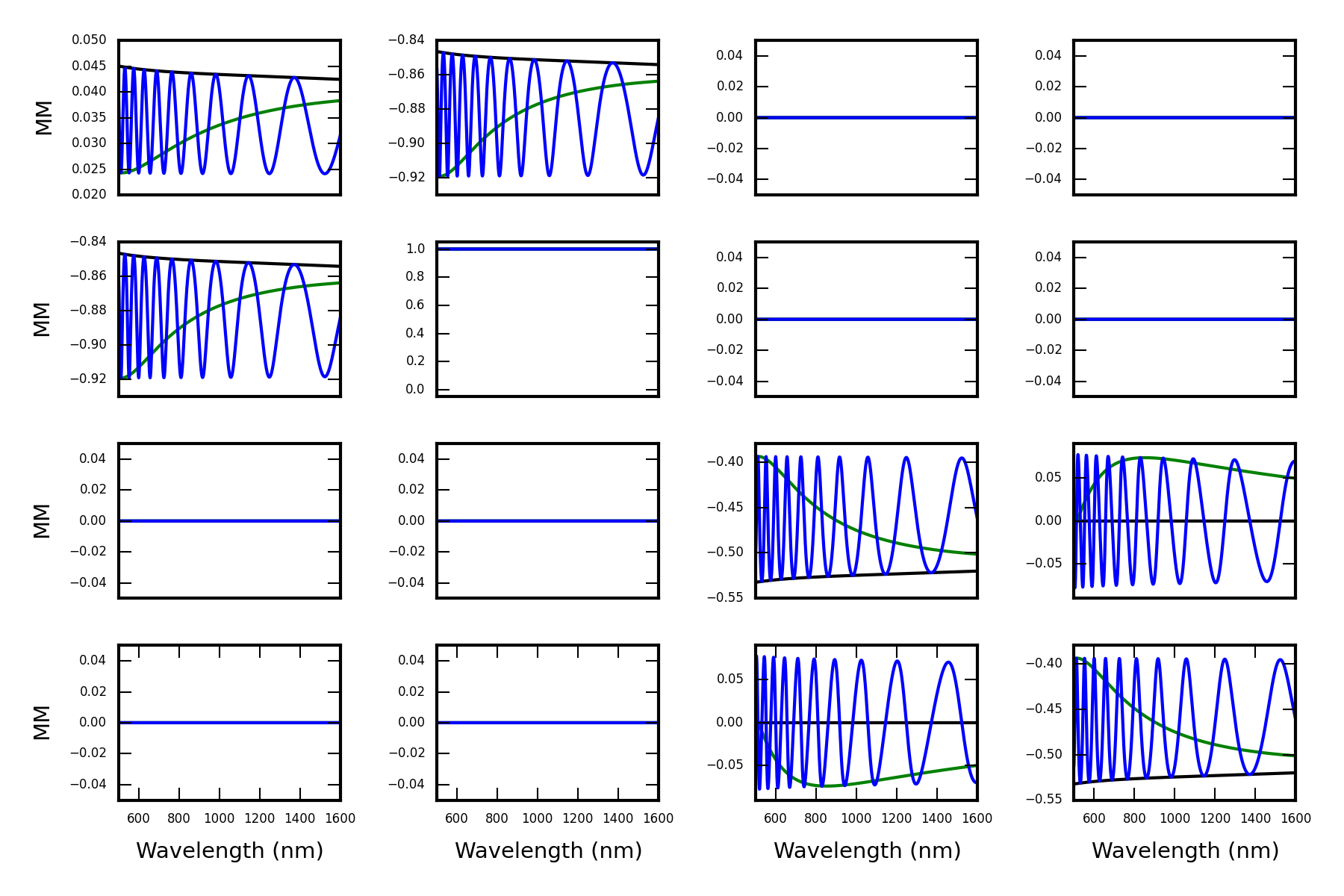}
}
\caption{The Mueller matrix of a reflection off a fused silica window at 45$^\circ$ incidence angle for various coating thicknesses.  The [0,0] element has been used to normalize all other matrix elements as described in equation \ref{eqn:MM_IntensNorm}. The black curve shows the Mueller matrix in reflection of uncoated isotropic bare fused silica.  The green curve shows the Mueller matrix for a reflection off the window including a quarter-wave thick isotropic MgF$_2$ coating at 600nm central wavelength (109nm physical thickness).  The blue curve shows the Mueller matrix for a window with a two-wave thick coating of isotropic MgF$_2$ with 2000nm central wavelength (2924nm physical thickness).  The [0,0] element shows the total uncoated reflection amplitude around 4.5\% as expected (black curve).  The standard quarter-wave MgF$_2$ anti-reflection coating (green) does indeed reduce the reflection amplitude to about 2.5\% at shorter wavelengths. As the incidence angle is 45$^\circ$, the central wavelength is slightly shifted and there is some retardance as seen in the $UV$ and $VU$ elements (green curve).  For the thick coating (blue), many cycles of constructive and destructive interference are seen as the coating passes integer multiples of quarter- or half- wave thicknesses.  }
\label{fig:thick_coated_window} 
\vspace{-6mm}
\end{center}
\end{figure}

The Berreman calculus results are shown for these three cases in Figure \ref{fig:thick_coated_window}.  As expected, the lower refractive index MgF$_2$ material does decrease the reflectivity when the thickness is roughly a quarter wave of optical path. This allows the back-reflection to be exactly out of phase with the incoming beam (180$^\circ$ phase of optical path) reducing the reflected intensity through destructive interference.  The blue curve shows the uncoated window has a reflectivity of roughly 4.5\% while the green curve shows a quarter-wave thick coating reduces the reflectivity to about 2.5\%.  The diattenuation terms $QI$ and $IQ$ similarly reduce from amplitudes of 0.04 to about 0.025.  The thick coating shows oscillations between these values as the interference is either constructive or destructive in multiples of quarter- or half-wave.

The $UV$ and $VU$ retardance terms also behave as expected in the presence of interfering waves of varying phase. Consider the simple approximation of the system interfering a single backwards traveling reflection off the fused-silica to MgF$_2$ interface with the incoming electro-magnetic wave.  This backwards traveling wave will have some coherent phase relation to the incoming wave through the thickness of the coating at that wavelength and incidence angle. Since the effective coating thickness in waves of phase is a function of wavelength, there will be variations in both amplitude and phase of the coherent interference with wavelength.  This beam will have varying intensities between S- and P- polarization directions (diattenuation) and it will arrive with varying phase against the incoming beam, introducing retardance.   

The addition of two sinusoids of the same frequency (wavelength) but differing phase can be expressed with a few simple terms using well known trigonometric functions. The squared amplitude goes as the sum square of the individual terms plus a coherence term based on the phase difference between interfering waves.  The square amplitude is:  $a_1^2 + a_2^2 + a_1 a_2 \cos(\delta)$ where $\delta$ is the phase difference between waves.  The phase of the summed wave is also dependent on the amplitudes and phase difference.  The new phase can be written as $\arctan( a_2 \sin(\delta),  a_1 + a_2 \cos(\delta))$.

\begin{wrapfigure}{r}{0.60\textwidth}
\centering
\vspace{-4mm}
\begin{tabular}{c} 
\hbox{
\hspace{-1.7em}
\includegraphics[height=6.6cm, angle=0]{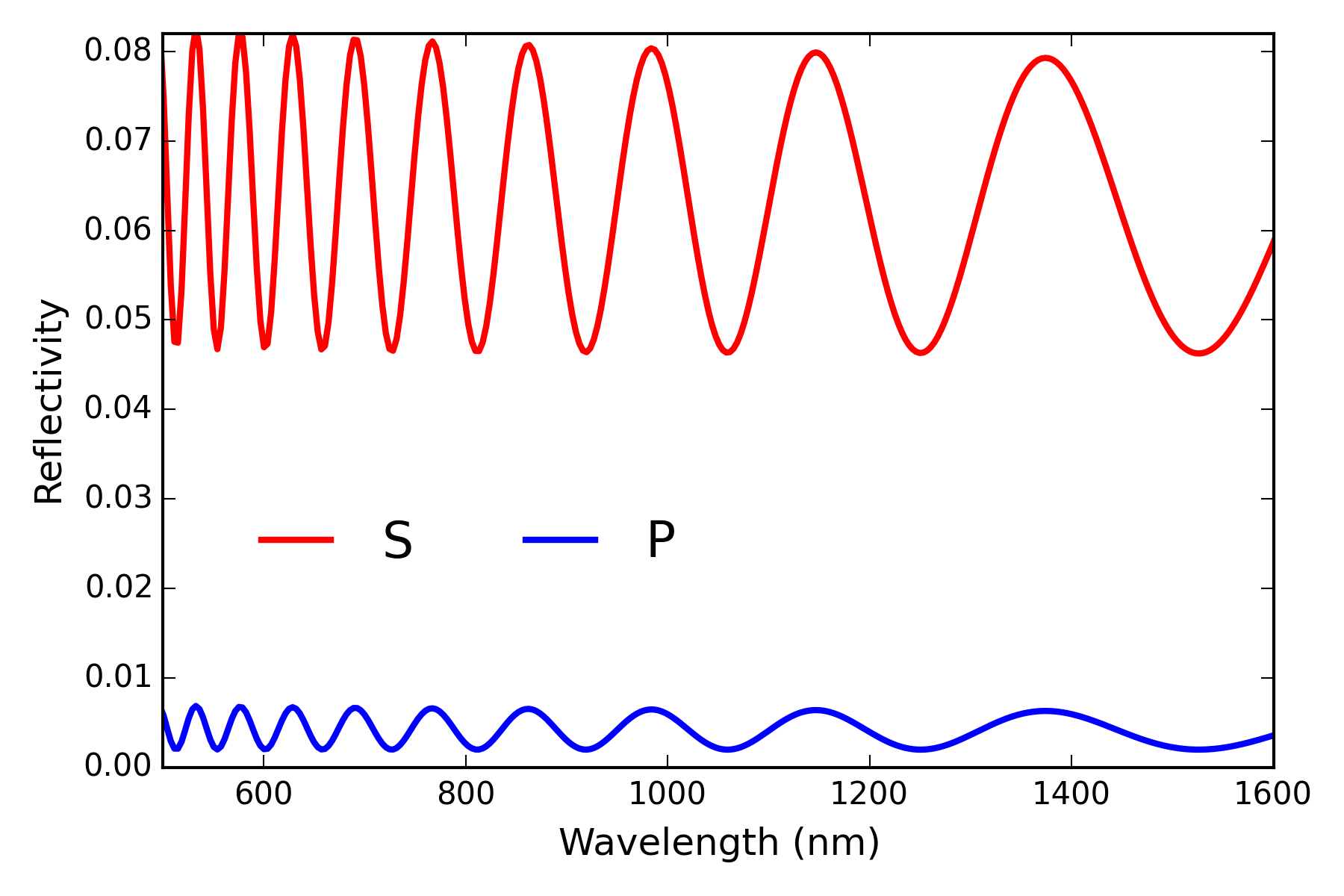}
}
\end{tabular}
\caption[Reflectivity for Thick Coated Window] 
{ \label{fig:normalized_parameters_thick_coated_window} 
The reflectivity for the thick isotropic MgF$_2$ coating on a fused silica window at 45$^\circ$ incidence angle. The reflectivity curves oscillate about their average values of 6.43\% for S- and 0.43\% for P-.   }
\vspace{-3mm}
 \end{wrapfigure}

The amplitude and phase of the summed waves represents the beam reflected off this window as it is approximated as the initial reflection off the MgF$_2$ surface plus the reflection off the MgF$_2$ to Fused Silica interface propagating back through the MgF$_2$ layer.  This simple approximation shows how even an isotropic material coated on an isotropic material can introduce retardance through coherent interference of waves.  In the Berreman calculus, the full problem is solved using transfer matrices based on the electric and magnetic fields which accounts for far more complexity than this simple description. 

Figure \ref{fig:normalized_parameters_thick_coated_window} shows the S- and P- reflectivity normalized by their average values for the thick MgF$_2$ coating along with retardance and diattenuation.  The oscillations in S- reflectivity have an amplitude of roughly $\pm$40\% and P- oscillates by over 20\% shown as the red and blue curves respectively. Given the coherent interference between forward and backward propagating fields, the retardance and diattenuation terms also oscillate with wavelength. Retardance derived from the phase difference between S- and P- waves shows oscillations of up to about 9$^\circ$. The diattenuation represents the amplitude ratio of S- and P- beams which also oscillates from 85\% to 92\%.

\subsection{Comparison with Zemax Coating Calculations}

Many readers will be familiar with the Abeles and Heavens calculus from thin film coating calculations. These are standard calculations included in common software such as TFCalc, Zemax or other industry standard modeling packages.  We used Zemax version 15.5 to verify that our Berreman calculation gives expected results in the limit of isotropic materials. In Zemax, we set up a coating that used similar refractive index values for isotropic {\it amorphous} MgF$_2$ coatings, a fused silica window and identical material thicknesses to compare our calculations. For those readers familiar with Zemax, the software has a coating analysis tool that uses the isotropic Abeles and Heavens matrices to compute the reflectivity, diattenuation and retardance for a specified layering of isotropic coating materials.  The Abeles and Heavens matrix formula can be considered the isotropic limit of the Berreman formalism where no birefringent materials are used in the calculation.

\begin{figure}[htbp]
\begin{center}
\vspace{-3mm}
\includegraphics[width=0.99\linewidth, angle=0]{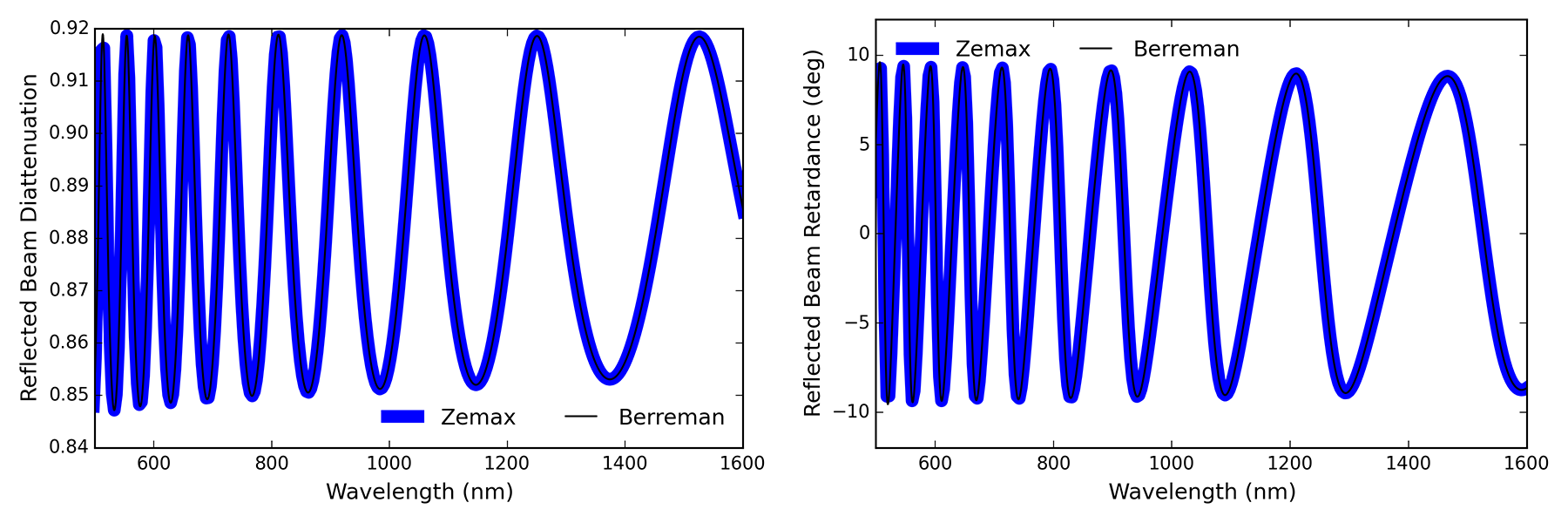}
\caption{The comparison of Zemax version 15.5 coating analysis tools and our Berreman calculations for a single thick layer of isotropic MgF$_2$ coated on a fused silica window illumiated at 45$^\circ$ incidence angle. The blue curves show the Zemax result for coating thickness of 2.0 waves at 2000nm primary wavelength. The black curve shows our derived values from the Berreman calculus code using similar values for the refractive index of the coating and window materials. The curves overlap completely showing good agreement between our code and the industry standard Zemax calculations. The left panel shows the diattenuation oscillating about an average value of $\sim$93\%. The right panel shows retardance oscillating about 0$^\circ$.  
 }
\label{fig:compare_zemax_berreman_dia_ret}
\vspace{-2mm}
\end{center}
\end{figure}

In Figure \ref{fig:compare_zemax_berreman_dia_ret}, we show how our software reproduces the familiar results from Zemax.  We show the diattenuation and retardance for the thick MgF$_2$ coating. Two waves thickness at 2000nm central wavelength gives 2924nm physical thickness of isotropic MgF$_2$ coated on top of fused silica at an indcidence angle of 45$^\circ$. The Zemax result is shown in blue while our calculations are shown in black. The curves overlap completely with differences of less than 0.01$^\circ$ retardance and 10$^{-5}$ diattenuation. For completeness, we also tested several other limiting cases.  Bare windows do indeed produce diattenuation but not retardance in both software packages as expected.  We tested several other material interfaces such as a coatings between various glass substrates. We find good agreement between Zemax and Berreman predictions in all cases. We also find similar levels of agreement between the thin film calculator program TFCalc and our calculations here. Likely these computational disagreements arise from how the Zemax specified coating refractive indices are interpolated to the wavelength of interest.  Both Zemax and our Berreman code adopt the same equation for refractive index of the fused silica substrate.

\subsection{Laboratory verification of intensity fringes in an isotropic substrate}

We verified these fringe amplitude and period predictions in the laboratory using a high resolution spectrometer and a window.  Meadowlark optics is fabricating the DKIST retarders and has several samples of the various window and crystal materials used in DKIST optics. To experimentally show the fringe periods and amplitudes, we derive here the fringe period and amplitude based on some simple physical assumptions outlined below. 

The fringes are coherent interference between the incoming wave and the back-reflected wave.  The back-reflected wave traverses the material twice. As such, the reflected beam sees an optical thickness of twice the physical thickness ($d$) times the refractive index ($n$).  We expect constructive interference when the window is integer multiples of half-wave thickness.  We expect destructive interference when the window is integer multiples of quarter-wave thickness (just like anti-reflection coatings).  To compute the fringe period, we solve for the wavelength interval required to change the optical path traversed by the back-reflected exactly one wave in Equation \ref{derive_fringe_period}.

\begin{wrapfigure}{l}{0.35\textwidth}
\centering
\vspace{-4mm}
\begin{equation}
\label{derive_fringe_period}
\frac{2dn}{\lambda_1} - \frac{2dn}{\lambda_2} = \frac{2dn(\lambda_1 - \lambda_2)}{\lambda_1 \lambda_2} \sim 1 
\end{equation}
\vspace{-8mm}
\end{wrapfigure}

For optics that are much thicker than a wavelength, we can approximate $\lambda_1$ and $\lambda_2$ as about the same.  For typical windows and crystals at visible wavelengths, the optic is at least millimeters thick and the fringe period is less than a part per thousand of the wavelength.  With this approximation, we can simplify Equation \ref{derive_fringe_period} to get a simple formula for the fringe period in wavelengths ($\delta\lambda$) expressed as the wavelength squared ($\lambda^2$) divided by the optical path for the back-reflected beam ($2dn$). 

\begin{wrapfigure}{r}{0.32\textwidth}
\centering
\vspace{-6mm}
\begin{equation}
\label{simplified_fringe_period}
\lambda_1 - {\lambda_2} = \delta \lambda = \frac {\lambda_1 \lambda_2} {2dn} \sim \frac{\lambda^2}{2dn}
\end{equation}
\vspace{-8mm}
\end{wrapfigure}

With this formula, it is straightforward to compute the expected fringe period as functions of the optical properties. As an example, a 1mm thick window with a refractive index of 1.5 observed at 500nm wavelength would see a fringe period of 0.083nm. Since the difference between wavelengths $\lambda_1$ and $\lambda_2$ is one part in 6,000, the assumption of $\lambda_1 \sim \lambda_2$ is quite reasonable.

\begin{figure}[htbp]
\begin{center}
\vspace{-3mm}
\includegraphics[width=0.995\linewidth, angle=0]{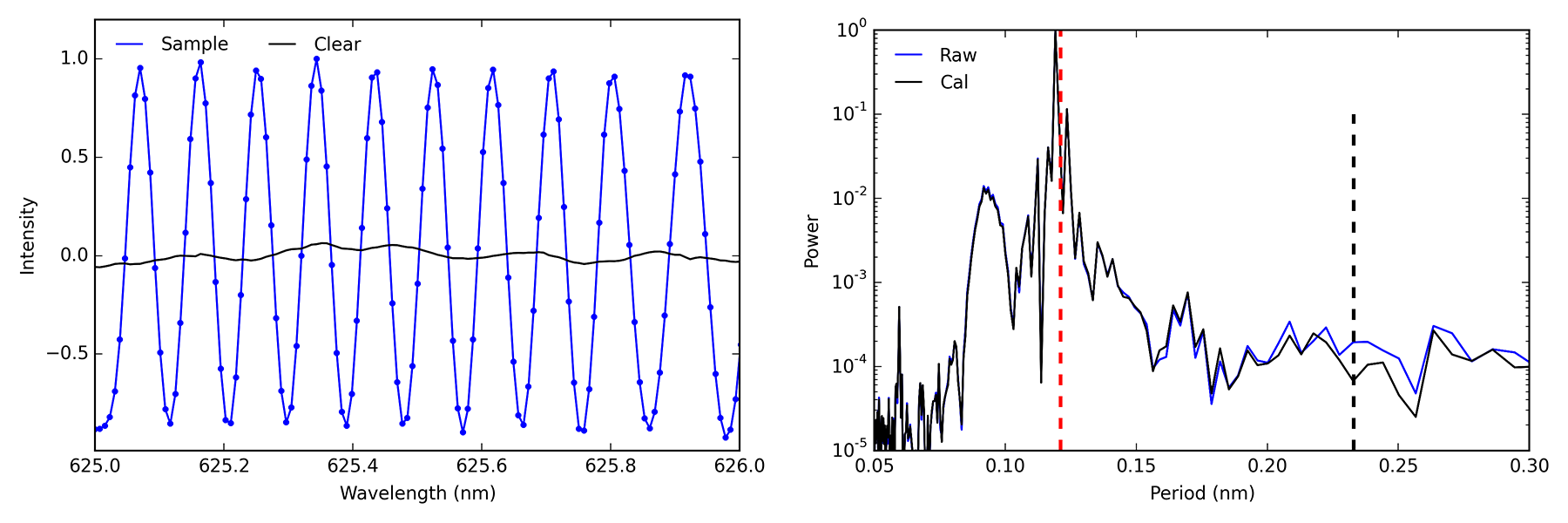}
\caption[Fringe Frequency for Infrasil] 
{ \label{fig:fringes_uncoated_window} 
The data (left) and power spectrum (right) for spectrophotometric data set.  The intensity is recorded for 1260 transmission measurements covering 625nm to 635nm wavelength using the Meadowlark SPEX 1401 spectrograph and a 1.1335mm thick Heraeus Infrasil 302 window. The power spectrum is computed as (FFT$^2$). For the left panel intensity data, strong oscillations of detected intensity are seen in blue with the baseline no-sample measurement in black.  There is some oscillation in the no-sample baseline data attributed to the PMT sensor window glass. However, the amplitude is much lower and the period of oscillation in this baseline is much longer than detected in the Infrasil window data set.  The right hand panel shows two power spectra. The blue curve shows the power in the raw intensity data uncalibrated for the baseline.  The black curve shows the power in the transmission spectrum of the sample scan calibrated by the no-sample scan. The vertical dashed red line shows the predicted fringe period from Equation \ref{simplified_fringe_period}. The black dashed line shows the period we find in the no-sample baseline data. The measured power is dominated by the single fringe period created by the Infrasil window which does not change through the calibration process. The fringe period agrees with the theory. The blue and black curves show minimal differences between raw data and baseline-calibrated transmission.  Other optics inside the spectrograph (e.g. detector window) do not contribute significantly to calculating window fringe periods. }
\vspace{-2mm}
\end{center}
\end{figure}

In the Meadowlark facility, a SPEX 1401 double grating 0.85-meter Czerny-Turner spectrometer is available with several sensor options. The light source is an Energetiq broad band fiber coupled plasma source using a 200 micron diameter core fiber. The fiber output is collimated to a $\sim$10mm beam by a Thor labs 90$^\circ$ fold angle silver-coated off-axis parabola mirror with an effective focal length of 15mm. The fiber light source and the OAP collimating mirror will produce some polarization expected to be at amplitudes less than a few percent at visible wavelengths. For this mounted OAP, the beam diameter is set by the exit of the housing after the mirror at an 11mm diameter. The mirror is oversized and mounted before this aperture, giving rise to a small field dependence of less than $\pm$0.4$^\circ$. Some spatial dependence on sampling the fiber core is introduced by having the system beam stop be a rectangular mask on the collimating mirror after the slit . 

At visible wavelengths, the measured resolution is $\frac{\lambda}{\delta\lambda}$ in the 25,000 to 45,000 range depending  primarily on how fully the grating is illuminated. The instrument profile of 0.016nm full-width half-maximum was measured with a neon discharge lamp at 653nm and the profile has Gaussian shape giving a resolving power of 40,800. Other lines at 585nm, 609nm,  633nm and 725nm with slightly different beam diameters gave resolving power in the range of 25,000 to 49,000 depending on grating illumination. 

The system is set up to nominally have a 10mm diameter collimated beam that is focused on the spectrograph entrance slit via a 50mm focal-length singlet. The fiber core is magnified by the 15mm to 50mm ratio and would theoretically fill a 0.67mm tall slit. The slit is 35 microns in width and is over 1mm high to pass the full fiber core image. The f/5 beam entering the spectrograph is stopped to f/7.8 beam by a rectangular aperture on the collimating mirror inside the spectrograph. The system uses photo-multiplier tubes to cover a range of wavelengths and delivers a typical measured signal to noise ratio around 1000. This noise level is dominated by systematic errors for integration times longer than 0.1 seconds. The baseline count rate was measured to vary by roughly 10\% in 200 minutes with a mostly linear trend.  

Meadowlark tested a window over the 625nm to 635nm wavelength range with the SPEX 1401 system. The window was 1.1335mm thickness of Heraeus Infrasil 302. Thickness was measured by a Haidenhain MT 60M metrology system with $\sim$0.5$\mu$m thickness accuracy. The measured transmitted wavefront error (TWE) at 632.8 nm wavelength is 0.021 waves peak-to-valley (PV) over an aperture of 12 mm. The beam deviation through the window is measured to be 0.26 arc-seconds over the same footprint.

A spectral sampling of 0.079nm was used to sample the beam at roughly two points per FWHM at the highest measured resolving power of R=45,000. We use 1 second per point integration for high signal-to-noise ratio at reasonable speed. Over the 10nm bandpass at 0.079nm sampling, we collected 1260 points. Scans were made with and without the sample to derive a transmission spectrum.  The beam was collimated and had a 10mm diameter footprint on the window. The transmission spectrum was dominated by a simple sinusoidal intensity fringe with a period of $\sim$0.12nm with transmission values oscillating between 87\% and 99\%. This measurement agrees very well with Equation \ref{simplified_fringe_period}. The Infrasil data sheet from the manufacturer (Heraeus) gives a refractive index of 1.457 at 632.8nm. We compute the period as $\frac{\lambda^2}{2dn}$ = 0.121nm.  

\begin{figure}[htbp]
\begin{center}
\vspace{-3mm}
\hbox{
\hspace{-1.0em}
\includegraphics[height=11.35cm, angle=0]{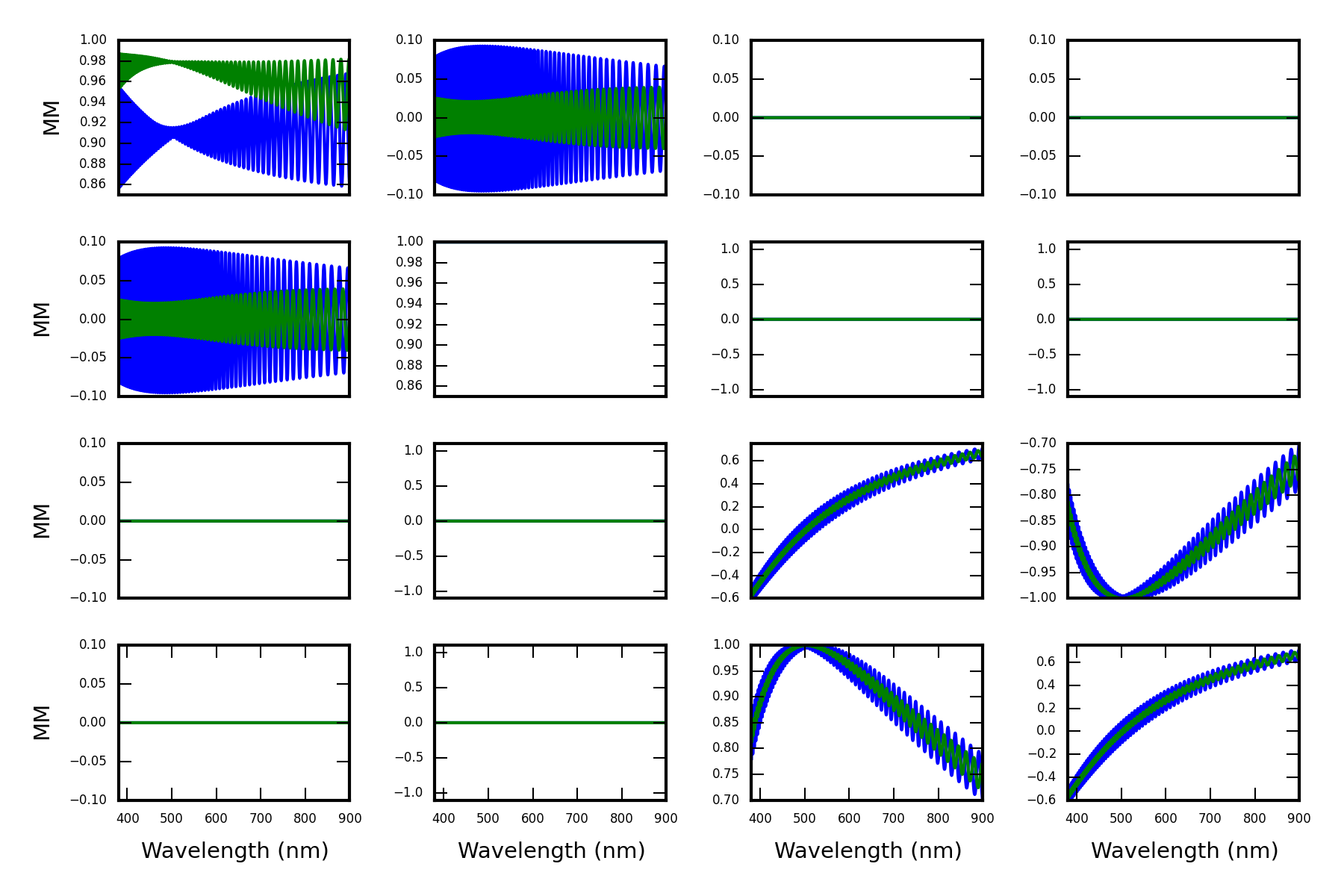}
}
\caption{The Mueller matrix of a zero-order quartz retarder from 380nm to 900nm wavelength. The quartz physical thickness is 13.587 microns to create a quarter-wave zero-order retarder at 500nm wavelength.  The blue curve shows the Mueller matrix for the uncoated quartz crystal at zero incidence angle.  The green curve shows the Mueller matrix for the optic after addition of 84.7nm of isotropic MgF$_2$ on both faces as an anti-reflection coating. All Mueller matrix elements besides [0,0] have been normalized by the [0,0] element for easy comparison as in Equation \ref{eqn:MM_IntensNorm}. The [1,1] element $QQ/II$ is identically 1 for all wavelengths. The nominal linear retardance is seen in the lower right hand 2x2 sub-matrix.  The transmittance and diattenuation (polarizance) can be seen in the upper left 2x2 sub-matrix. The fringes form an envelope in both sub-matrices showing impact for transmittance, diattenuation and retardance. See text for details. }
\label{fig:quartz_ZWP_Comparison} 
\vspace{-6mm}
\end{center}
\end{figure}

The 10nm bandpass records roughly 84 fringe periods allowing us to do a simple Fourier analysis. Figure \ref{fig:fringes_uncoated_window} shows the power spectrum of the measurements. The blue curve shows the power in the raw detected intensity data without calibration by a baseline. The black curve shows the power in the transmission spectrum computed after calibration of the data using the no-sample scan. The vertical dashed red line shows the predicted fringe frequency from Equation \ref{simplified_fringe_period} which agrees with the measured peak in both raw counts and calibrated transmission.  The agreement between blue and black curves as well as the absence of significant power at other periods shows that the detector window internal to the spectrograph does not contribute to the fringes through the calibration process.  

To compute the likely fringe amplitude, the expected electric field amplitudes must be added or subtracted coherently.  As a first-order approximation for highly transmissive optics, we can simply consider the reflection from the back surface interfering coherently with the front surface. The initial reflection has an intensity of 3.5\% at the interface between air and Infrasil.  The transmitted beam reflects off the back surface and then transmits again through the front surface with an intensity that scales with the back surface reflectivity twice multiplied by the transmission of air to Infrasil. This is roughly 96.5\% (3.5\% $\times$ 96.5\%) $\sim$ 3.38\%.

\begin{wrapfigure}{l}{0.3\textwidth}
\centering
\vspace{-7mm}
\begin{equation}
\label{add_efields_via_intensity}
I_{tot} = (\sqrt{I_{front}}  \pm  \sqrt{I_{back}})^2
\end{equation}
\vspace{-7mm}
\end{wrapfigure}

Electric field amplitudes go as the square root of the intensity.  With this assumption of a single reflected beam interfering, we can predict the total intensity as constructive and destructive interference at each wavelength. We can compute the intensity interference amplitude using the predicted reflection and transmission coefficients after converting to field amplitudes as in Equation \ref{add_efields_via_intensity}.  For our nominal 3.5\% front surface reflection interfering with a 3.3\% back surface contribution, we expect to see the transmission oscillate from 86\% to 99.9\% for complete destructive and constructive interference.  We detected transmission (intensity) fringes from 87\% to 99\%, which we consider to be a good experimental match.  Detected fringe amplitude can be decreased by reduced spectral resolution, optic wavefront flatness, optical wedge and several instrumental degradations. Simple single-reflection theory predicts 14\% intensity variation and detect 12\% variation

\section{Quartz Retarders \& Coatings}

In this section, we outline comparisons between the MHW implementation of the Berreman calculus and predictions from several others in the astronomical literature about polarization fringes.  The updated 2015 MHW textbook has a chapter on crystals and achromatic retarders that shows how the formalism can be used in predicting fringes as well as basic polarization properties of the retarder.

Clarke, Semel, Aitken, Hough and others had developed some specific frameworks for predicting fringes in astronomical settings but the Berreman calculus is the general case which can handle all orientations of many-layered birefringent media with arbitrary material orientations in the layers \cite{2003A&A...401....1S, 2004JOptA...6.1041C, Clarke:2004gm,2005A&A...434..377C, Clarke:2009ty, 2004JOptA...6.1047C, Aitken:2001ih}

A true zero-order quartz retarder is incredibly difficult to manufacture. The thickness is less than 50$\mu$m for zero-order retardance at visible wavelengths. At 525nm wavelength, the computed thickness of the crystal using the CVI Melles Griot catalog formulas for refractive index and birefringence is 13.587 microns. The refractive indices are (1.556 and 1.547) for extraordinary and ordinary beams. The optical thickness is thus (40.28, 40.03) waves and the difference between e- and o- beams is the design retardance of 0.25 waves. It is straightforward to create the Berreman model of such an un-manufacturable part for comparison with the theoretical predictions of Clarke 2004 and 2005 \cite{2004JOptA...6.1041C, Clarke:2004gm,2005A&A...434..377C, Clarke:2009ty, 2004JOptA...6.1047C}.

\begin{wrapfigure}{l}{0.55\textwidth}
\centering
\vspace{-4mm}
\begin{tabular}{c} 
\hbox{
\hspace{-1.5em}
\includegraphics[height=6.3cm, angle=0]{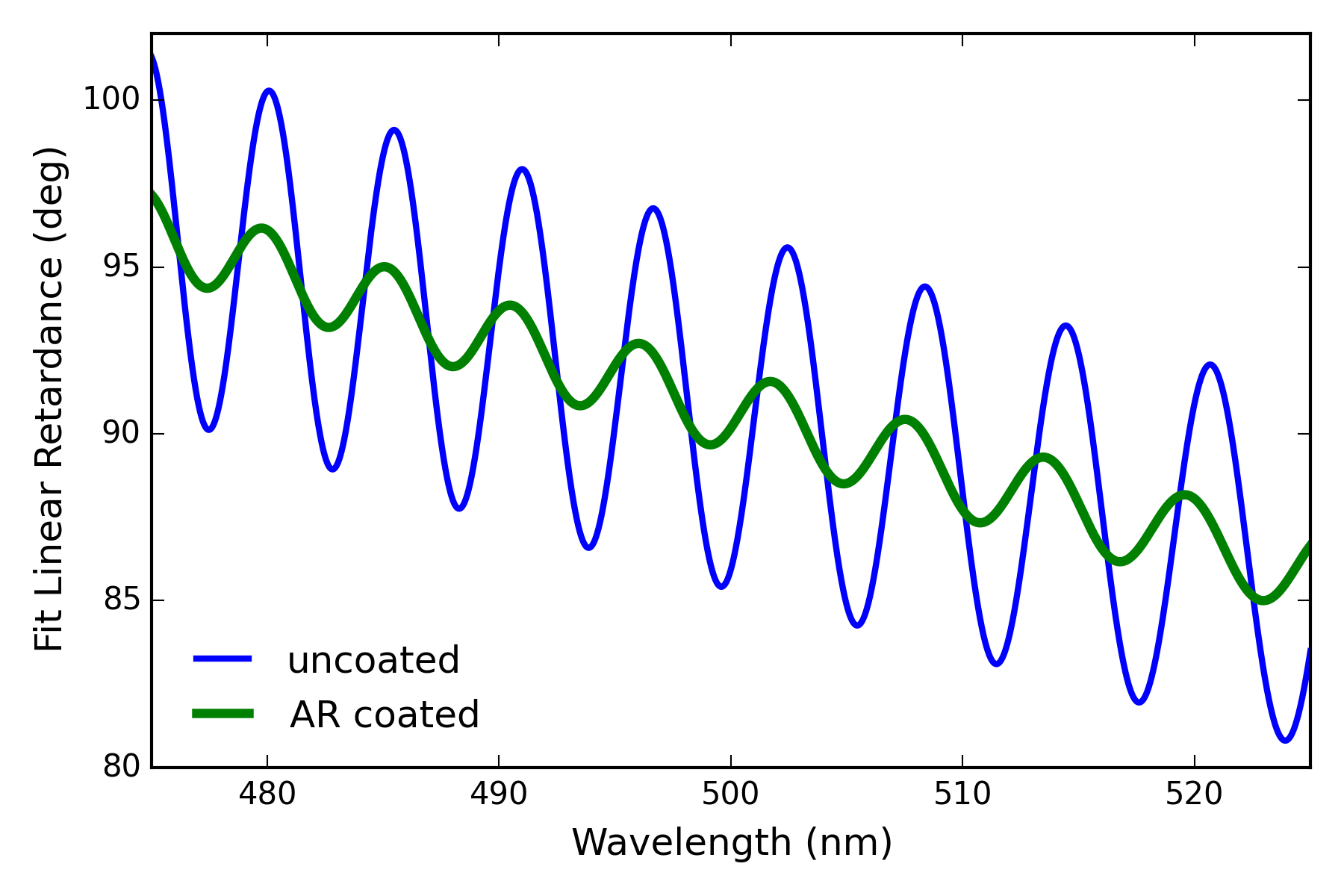}
}
\end{tabular}
\caption[Retardance for Zero-order Quartz Retarder] 
{ \label{fig:zero_order_quartz_linear_retardance_fit} 
The linear retardance fit to a zero-order quartz crystal retarder of 13.587 microns physical thickness.  Blue shows uncoated. Green shows anti-reflection coated.  The blue curve can be compared with Clarke 2004. \cite{Clarke:2004gm} We use the Sueoka \cite{Sueoka:2016vo} revised birefringence measurements, and we get a slightly different retardance than when using the CVI refractive indices giving 13.587$\mu$m as the exact zero-order retardance thickness for 500nm wavelength. }
\vspace{-4mm}
 \end{wrapfigure}

Figure \ref{fig:quartz_ZWP_Comparison} shows the Mueller matrix for this zero order retarder derived using our Berreman calculus scripts. We ran the computation at a spectral sampling of $\lambda$/$\delta\lambda$ of 2,000,000 to fully resolve all possible fringes. The blue curve shows an uncoated quartz crystal with a [0,0] transmittance term ($II$) of around 91\%. This amplitude is expected for two Fresnel reflection losses at interfaces of refractive index $\sim$1.5. The green curve shows the impact of anti-reflection coatings. The coating is on both exit and entrance interfaces with 84.7nm of isotropic MgF$_2$ which is about a quarter-wave at wavelengths around 525nm. The transmittance of the anti-reflection coated (green) model shows transmission is around 98\%.

The retardance at 500nm wavelength is essentially a quarter-wave as can be seen in the lower right hand quadrant of the Mueller matrix in Figure \ref{fig:quartz_ZWP_Comparison}.  The $UV$ and $VU$ terms have amplitudes near 1 with $UU$ and $VV$ terms with amplitudes near zero at 525nm wavelength as expected for a quarter-wave linear retarder. For the two reflections of the uncoated part (blue curve), the diattenuation seen in $IQ$ and $QI$ terms reaches ampliutdes of nearly 10\%.

In Figure \ref{fig:zero_order_quartz_linear_retardance_fit} we show a fit to the linear retardance for the Figure \ref{fig:quartz_ZWP_Comparison} Mueller matrices.  Even though the diattenuation does couple into the $UV$ quadrant terms, the linear retardance fit still gives the expected result.  The part is a quarter-wave linear retarder at 500nm, thought with substantial fringes. The blue curve can be compared with Clarke 2004 \cite{Clarke:2004gm}

\begin{figure}[htbp]
\begin{center}
\vspace{-0mm}
\includegraphics[width=0.99\linewidth, angle=0]{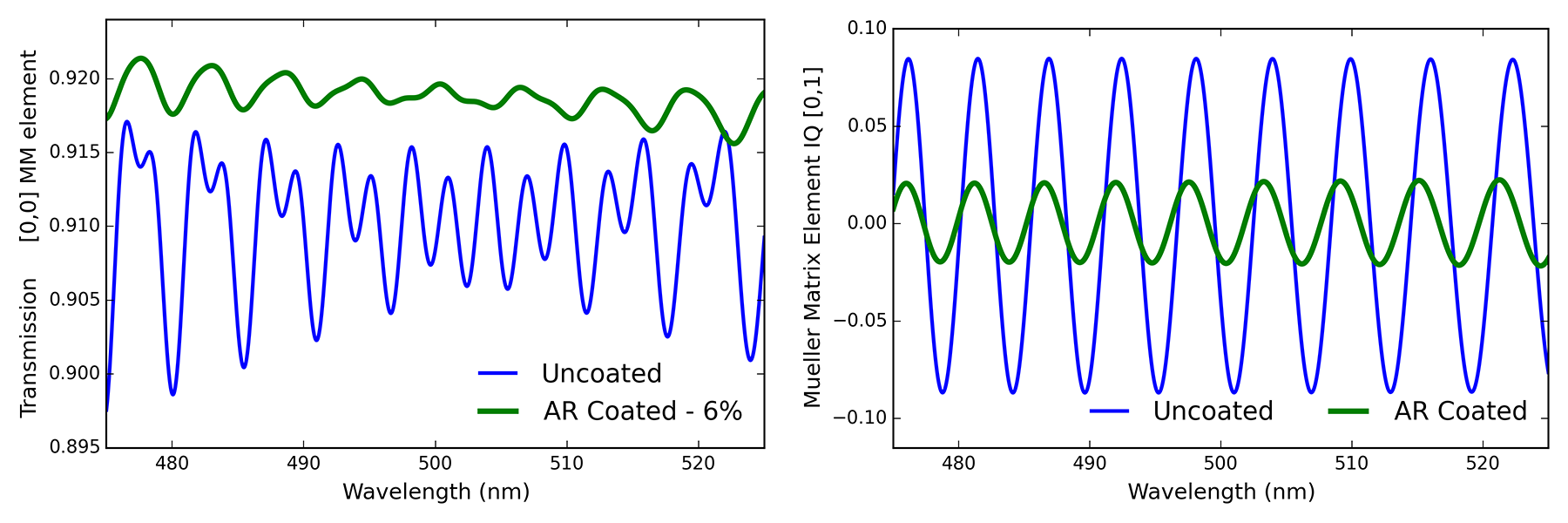}
\caption{The left panel shows transmittance to a zero-order quartz crystal retarder of 13.587 microns physical thickness.  Blue shows uncoated. Green shows the anti-reflection coated part with 6\% reflection subtracted for clarity.  The AR coated transmittance can be seen in the Mueller matrix of Figure \ref{fig:quartz_ZWP_Comparison} and is around 98\% at 525nm wavelength.  The right panel shows the polarizance again with blue as uncoated, green as anti-reflection coated with a quarter wave of isotropic MgF$_2$ using a coating central wavelength of 525nm for 84.7nm physical thickness. The blue curves can be compared with Clarke 2004. \cite{Clarke:2004gm} Slight wavelength shifts will also be seen as we use the Sueoka \cite{Sueoka:2016vo} revised birefringence formulas.}
\label{fig:zero_order_quartz_transmittance_polarizance}
\vspace{-2mm}
\end{center}
\end{figure}

Similar comparisons can be made for transmittance and polarizance.  Figure \ref{fig:zero_order_quartz_transmittance_polarizance} shows the same wavelength region as Clarke 2004 \cite{Clarke:2004gm}.  For transmittance, the fringe pattern has multiple components due to the birefringence of the medium. We have done similar comparisons for a true zero-order half wave retarder and we reproduce the results of Clarke for the 27$\mu$m thick quartz crystal. The Berreman calculus is a robust framework that can be used not only in the limit of parallel plates and perfectly cut crystals In further sections, we will model thicker multi-order crystals, misaligned crystals and compare with measurements of 2mm to $>$13mm multi-crystal stacked retarders.

\subsection{Multi-order Quartz Retarder Data \& Models}

We show a Berreman model for a multi-order quartz crystal retarder compared to high spectral resolution scans with the Meadowlark Spex instrument. The retarder thickness was measured to have 575.4 $\pm$0.5 microns thickness and was mounted for testing with retarder fast and slow axes at 45 degrees to the grating rulings. This thickness corresponds to about 16.5 waves of retardance at 630nm wavelength. The TWE for the crystal is 0.034 waves PV at 632.8 nm wavelength over a clear aperture of 12 mm. Beam deviation was measured to be 0.21 arc-seconds. 

The retarder was mounted in a collimated beam and baseline scans without the sample were also recorded. A slight blemish in the beam attributed to the fiber was identified and the beam was stopped to roughly f/10 inside the spectrograph to ensure this blemish did not impact the data. This beam underfilled the grating and likely had reduced spectral resolving power.  In the data sets presented in this section, we increased the spectral sampling to cover 30nm bandpass at steps of 2pm giving an effective sampling at $\frac{\lambda}{\delta\lambda}$ of about 315,000.  We sample the 16pm FWHM beam with about 8 points with a total of roughly 15000 points over the bandpass. 

A Fourier analysis of the data set give a fringe period of 0.22nm.  The power spectrum is dominated by a single somewhat broad peak at 0.22nm without any other significant features in the 0.05nm to 0.5nm period range.  The theoretical period of $\lambda^2$/2dn is 0.2226nm for the extraordinary beam of refractive index 1.551 and 0.2239nm for the ordinary beam at a refractive index of 1.543. 

\begin{figure}[htbp]
\begin{center}
\vspace{-3mm}
\hbox{
\hspace{-1.0em}
\includegraphics[height=5.7cm, angle=0]{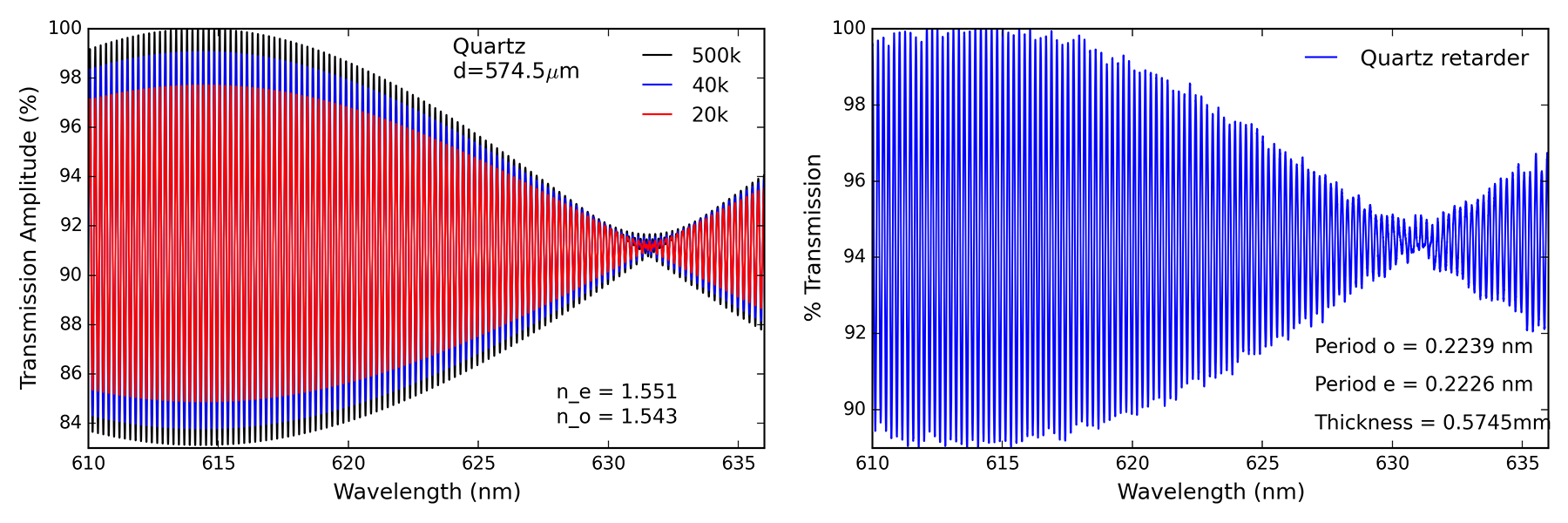}
}
\caption[] 
{  
The left panel shows transmission fringes for the Quartz retarder computed using our Berreman code over the bandpass 610nm to 650nm. Black shows the Berreman calculation at spectral sampling of 500,000.  Blue and red show the fringes convolved with a Gaussian instrument profile of resolving power 40,000 and 20,000 respectively. The right panel shows the Meadowlark Spex 1401 data set.  Transmission ranges from 89\% to 100\% for a fringe amplitude of 11\%, consistent with the Berreman calculations at R=20,000 resolving power.
\label{fig:quartz_fringes_berreman}  }
\vspace{-6mm}
\end{center}
\end{figure}

We created fringe predictions with our Berreman code as shown in Figure \ref{fig:quartz_fringes_berreman}. The fringes were derived at spectral sampling of $\delta\lambda/\lambda$ = 500,000 and infinite spectral resolving power.  To show the impact of instrument resolving power on fringe amplitude, we convolve these Berreman predictions with Gaussian profiles of varying widths corresponding to the Spex range of R=20,000 to R=40,000.  

The higher refractive index Quartz crystal has transmission ranging from 99.99\% to 81.8\% for a fringe amplitude around 18\% using the simple single-reflection analytic equation. When convolving this theoretical curve with a Gaussian profile at resolving power of R=20,000 the amplitude decreases to about 12\%. In addition, the interference between the extraordinary and ordinary beams gives rise to a much slower amplitude modulation at a period of roughly 35nm at 630nm wavelength.  The measurements show the minimum fringe amplitude clearly around 631nm wavelength in Figure \ref{fig:quartz_fringes_berreman} with fringe amplitudes rising quickly to shorter and longer wavelengths. This is very similar to our Berreman calculation of  Figure \ref{fig:quartz_fringes_berreman}.

\section{Bicrystalline Achromats \& SPINOR at the Dunn Solar Telescope}

In this section, we outline bi-crystalline retarders and compare with three retarders in use at a solar spectropolarimeter.  Bi-crystalline retarders have more complex fringe patterns because there is an internal interface between birefringent media. Typical bi-crystalline retarders have one crystal {\it subtract} retardance from the other to compensate for chromatic variation of birefringence. Sometimes the two crystals are different materials. Many designs include quartz, MgF$_2$ and / or sapphire in various arrangements. These crystals often are either optically contacted, coated with a many-layer anti-reflection coating and sometimes spaced with refractive index matching oil or air.  Wedges between the crystals are not uncommon, complicating the modeling of the fringes and crystal spacing. With a reflection propagating back through the optic from multiple internal interfaces, the birefringence of the crystal in addition to diattenuation complicates the Mueller matrix of the fringes.

Given that quartz is positive uniaxial while sapphire is negative uniaxial, a common arrangement is to cut the crystals as retarders and then mount the two plates oriented such that their crystal axes align. The pairing of positive and negative crystals will cause nearly complete {\it subtraction} of retardance of each plate to form a net retardance near the desired value with broad wavelength coverage.  Typically, a small thickness adjustment of a few microns is performed via polishing to one or both crystals to exactly match 0.5 waves measured retardance at 633.443 nm wavelength to form the bicrystalline achromatic design. In our code, we adjust the sapphire crystal analytically to match the retardance given our analytical refractive index formulas.

\begin{wrapfigure}{l}{0.55\textwidth}
\centering
\vspace{-3mm}
\begin{tabular}{c} 
\hbox{
\hspace{-1.1em}
\includegraphics[height=6.4cm, angle=0]{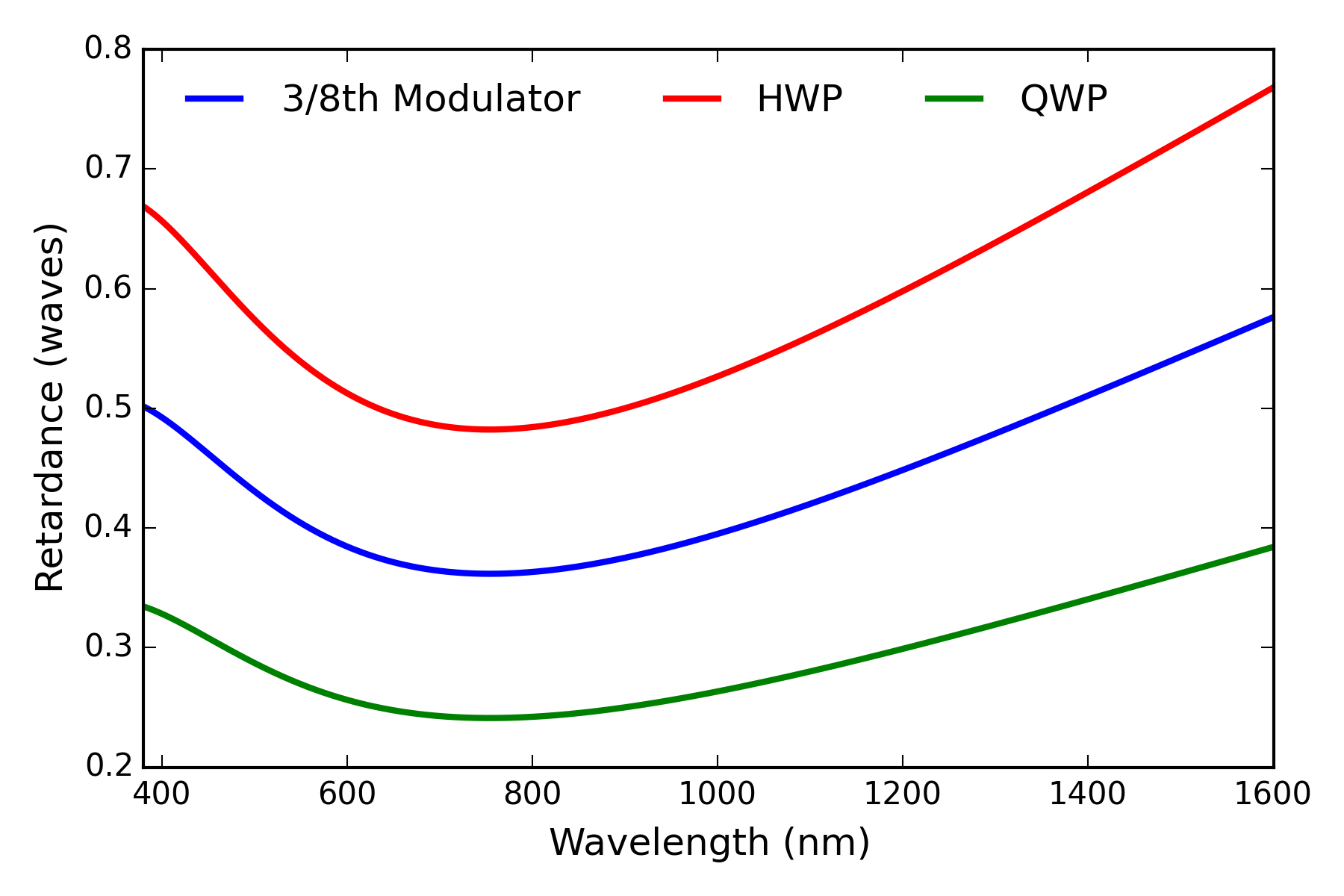}
}
\end{tabular}
\caption[Retardance of Designs for SPINOR 1/4, 3/8 and 1/2 wave] 
{ \label{fig:spinor_retarders_bestfit} 
The linear retardance predicted for the three SPINOR bi-crystalline achromats. SPINOR retarder designs at 633nm wavelength are to 1/4 wave, 3/8 wave and 1/2 wave retardance. The 1/4 wave is shown in green with crystals at (0.960209mm, 1.098837mm) thickness. The 3/8 wave retarder is shown in blue with crystals of (1.438883mm, 1.648256mm) thickness. The 1/2 wave retarder is shown in red with crystals at (1.920417mm, 2.197675mm) thickness.  }
\vspace{-4mm}
 \end{wrapfigure}

The Spectro-Polarimeter for Infrared and Optical Regions (SPINOR) is a spectropolarimeter at the Dunn Solar Telescope in Sacramento Peak, New Mexico operated by the National Solar Observatory \cite{2006SoPh..235...55S}.  This instrument is capable of polarimetry from 400 nm to 1600 nm and can be operated with up to four separate cameras. This allows for the simultaneous observation of several visible and infrared spectral regions with full Stokes polarimetry. The layout of the optics on a bench is determined for each observing run, giving the system flexibility to observe a wide combination of spectral lines.

\begin{wrapfigure}{r}{0.50\textwidth}
\centering
\vspace{-0mm}
\begin{tabular}{c} 
\hbox{
\hspace{-1.5em}
\includegraphics[height=6.2cm, angle=0]{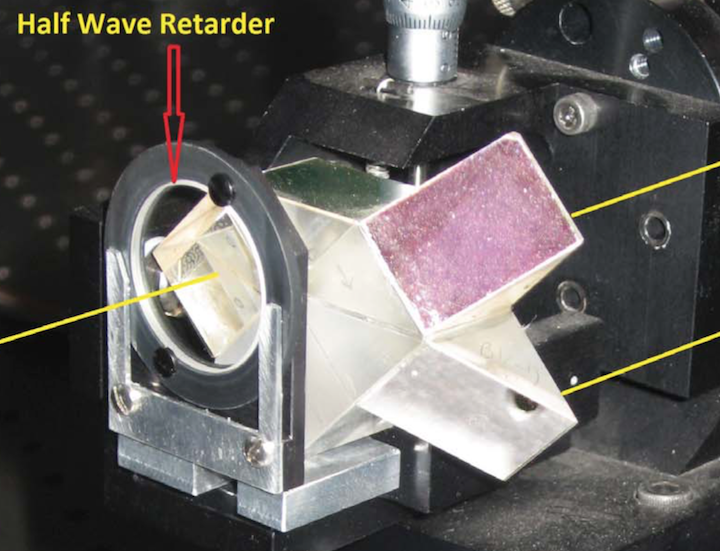}
}
\end{tabular}
\caption[Half-wave Retarder in SPINOR PBS] 
{ \label{fig:spinor_halfwave_on_PBS} 
The half-wave retarder mounted with the SPINOR polarizing beam splitter near the slit. Courtesy Doug Gilliam, DST staff.}
\vspace{-3mm}
 \end{wrapfigure}

For SPINOR, there are three retarders for the instrument.  There is a 3/8 wave sapphire-quartz bicrystalline achromat that is used as the modulator.  There is a 1/4 wave retarder used for calibration.  A 1/2 wave retarder is mounted just in front of the polarizing beam splitter.  We have some nominal information on the design of these parts.  Figure \ref{fig:spinor_retarders_bestfit} shows our best-fit retardance curves for these three optics. Figure \ref{fig:spinor_halfwave_on_PBS} shows the half-wave retarder mounted immediately ahead of the polarizing beam splitter.

The DST staff supplied the 1/4 retarder nominal design retardance values. For the quartz crystal, the design is 13.7446 waves retardance and -13.980956 waves retardance for the sapphire crystal at the design wavelength of 633.443 nm.  The refractive indices at the design wavelength using our formulas for quartz gives (1.5426, 1.5517) for $(n_o, n_e)$ and sapphire (1.7659, 1.7578). With these refractive indices, the specified design retardance values set the part thickness as the design retardance multiplied by the design wavelength divided by the birefringence $(n_e - n_o)$.  This equation gives 0.960mm thickness for quartz and 1.099 mm thickness for sapphire.

Our engineering efforts are possibly in error as the formulas used years ago by Meadowlark Optics to create the design are unknown. The design plate retardance in waves also slightly deviates from a quarter wave retarder when using the standard refractive index formulas from vendor catalog or recent analysis \cite{Sueoka:2016vo}.  

With these design values, quartz crystal in the quarter-wave retarder has (o,e) beam thicknesses of (2338.39, 2352.14) waves.  The sapphire has (o,e) thicknesses of (3063.27, 3049.29) waves. When ordinary and extraordinary rays are summed, the part thickness is (o,e) = (5401.667, 5401.430) which is 0.23 waves retardance. A quarter wave of retardance in crystal quartz corresponds to $\sim$14 microns of crystal optical path. Polishing to remove only a few microns thickness would be enough to adjust the retardance to exactly quarter-wave. The thickness change for 0.2 waves of retardance is 0.95 microns of quartz crystal material. The vendor would have polished the crystals to an approximate thickness and then adjusted one or the other crystal to match a design retardance. This is one outstanding uncertainty. 

\begin{wraptable}{l}{0.67\textwidth}
\vspace{-2mm}
\caption{SPINOR Retarder Physical \& Optical Properties 633.443nm}
\label{table:SPINOR_Retarders}
\centering
\begin{tabular}{l l l l l l l}
\hline\hline
$\delta$	& Qtz		& Qtz		& Qtz-O		& Sapph.		& Sap  		& Sap-O	\\
waves	& $\mu$m		& biref.		& waves		& $\mu$m		& biref.		& waves	\\
\hline
\hline
0.250 	& 959.222 	& 14.11153 	& 2349.81		& 1098.873	& -14.2986	& 3049.28		\\
0.375	& 1438.883	& 21.16729	& 3524.71		& 1648.256 	& -21.4479	& 4573.92 	\\
0.500	& 1918.511	& 28.22305	& 4699.61		& 2197.675 	& -28.5972	& 6098.56 	\\
\hline
\hline
\end{tabular}
The properties of the bicrystalline achromatic retarders used for SPINOR. The columns are: design retardance in waves, quartz physical thickness ($\mu$m), quartz crystal birefringence in waves, optical thickness in waves for the ordinary ray and then the same for the sapphire crystal.
\vspace{-3mm}
\end{wraptable}

The 3/8 wave modulator for SPINOR was designed to be scaled from the 1/4 wave retarder. The quartz crystal has 21.16729 waves retardance. The sapphire crystal has -21.4479 waves retardance at 633.443 nm wavelength.  We compile some of the optical and mechanical properties of the three SPINOR retarders in Table \ref{table:SPINOR_Retarders}.  The designs we use for our fringe analysis could be further optimized if we had high accuracy retardance measurements over a broad wavelength.  However, we find that the values in Table \ref{table:SPINOR_Retarders} are sufficient for comparing our fringe predictions with observations using SPINOR and the retarders at the DST.

\subsection{Air Gap \& Coating Impact on Fringe Amplitude and Spectral Frequency}

For the analysis we present, we model the air gap as 20 microns. The gap is set by spacer spheres placed on the outer edge of the aperture. On one side of the aperture, the spacers are slightly thicker. This results in a slight tilt to the second optic.  This does reduce ghosting for the modulator and half-wave retarders near focal planes. However, the small angle has minimal impact on the polarization fringes reported here as the beam overlap for coherent interference is not substantially changed.

The anti-reflection coating is almost certainly a many-layer broad-band AR coating of unknown formulation. The company ZC\&R had developed a coating to nominally cover the 450 nm to 1600 nm wavelength range. According to the DST staff, the coating should have roughly 1\% reflectivity from 450 nm to 650 nm wavelength and 0.5\% out to 1600 nm wavelength with some slow spectral oscillations. We have seen no records of the measured coating performance and will consider the impact of coating variation in later sections.

\begin{wrapfigure}{l}{0.65\textwidth}
\centering
\vspace{-4mm}
\begin{tabular}{c} 
\hbox{
\hspace{-1.5em}
\includegraphics[height=7.5cm, angle=0]{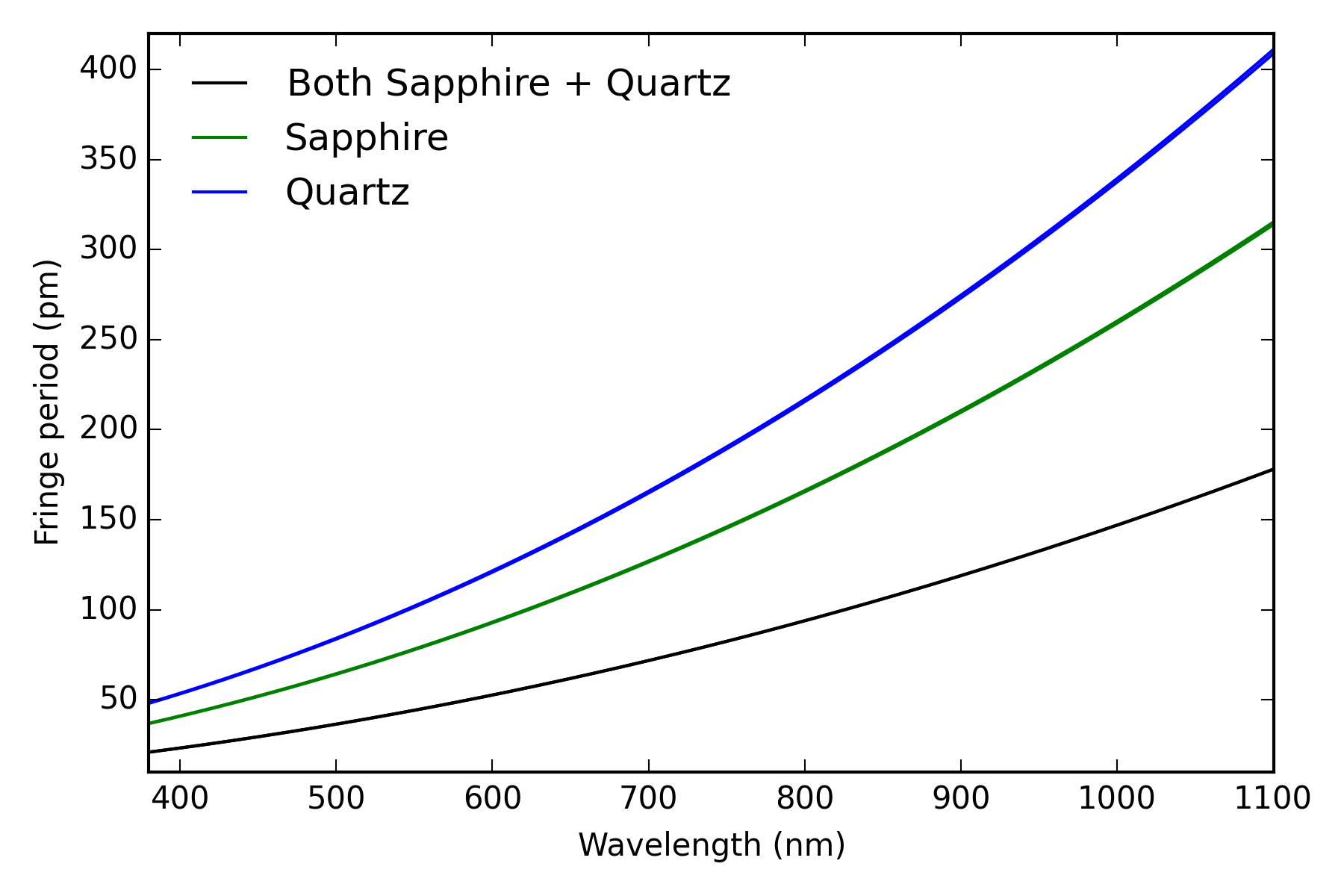}
}
\end{tabular}
\caption[Retardance for 3/8 Wave SPINOR Modulator] 
{ \label{fig:spinor_fringe_period} 
The fringe period predicted using the optical thickness of the quarter-wave retarder and Equation \ref{simplified_fringe_period} for period  ($\lambda^2$/2dn).  The o- and e- beam fringe periods are both plotted, but are within the thickness of the plotted line. Blue shows the fringe through the quartz crystal, green shows the sapphire crystal, black shows both crystals together as a bi-crystalline achromat.  See text for details. }
\vspace{-2mm}
 \end{wrapfigure}

For simplicity, we use a single-layer quarter-wave of isotropic MgF$_2$ with a central wavelength near 600 nm to illustrate how the fringe amplitude is reduced when using AR coatings on this part. Though even isotropic materials can produce retardance and diattenuation when used at non-zero incidence, we are not considering incidence angle variation in this section.  We proceed with an analysis of the on-axis beam only using this coating to illustrate how the SPINOR fringes are dominated by the internal reflections and birefringence of the two crystals.

For our analysis of thick crystals, the concept of a fringe spectral period is reasonably well defined.  As shown in Table \ref{table:SPINOR_Retarders}, our crystals are typically several thousand waves of optical path thickness. The coherent interference with wavelength through multiples of quarter- and half- wave is dominated by the change in wavelength.  Typical crystals of $>$1 mm thickness are thousands of waves optical path thickness at visible wavelengths. This optical thickness is computed as the physical thickness times the refractive index divided by the wavelength (d*n/$\lambda$).  With an essentially constant refractive index over a narrow bandpass, the wavelength changes by one wavelength divided by the optical path and the interference will change from constructive to destructive.  This allows us to apply simple Fourier analysis to the fringes for thick crystals and to speak of the {\it spectral period} of the fringe as in Equation \ref{simplified_fringe_period}. 

\begin{wrapfigure}{r}{0.65\textwidth}
\centering
\vspace{-4mm}
\begin{tabular}{c} 
\hbox{
\hspace{-0.9em}
\includegraphics[height=7.5cm, angle=0]{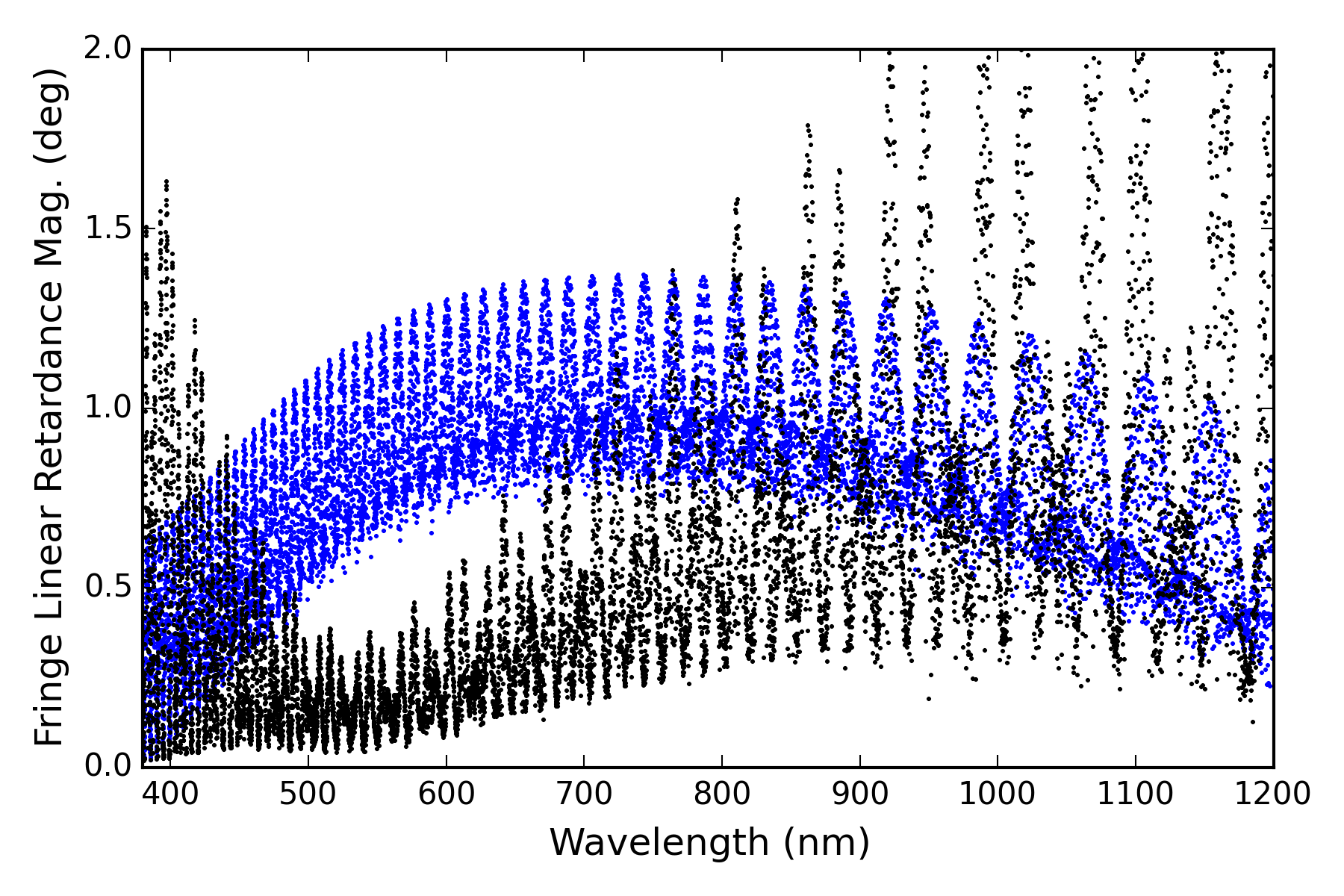}
}
\end{tabular}
\caption[Retardance for 3/8 Wave SPINOR Modulator] 
{ \label{fig:spinor_modulator_linear_retardance_fringe} 
The amplitude of the linear retardance fringes in narrow spectral bandpasses of R=10,000 spectral sampling for the SPINOR 3/8 wave modulator. Black is the linear retardance fringe magnitude using a 20$\mu$m air gap and a single-layer anti-reflection coating.  Blue shows an uncoated model with crystals optically contacted. See text for details.  }
\vspace{-5mm}
 \end{wrapfigure}

As an example, the SPINOR modulator at 400nm has quartz crystal thicknesses of (5637.9, 5603.5) waves for the e- and o- beams and sapphire crystal thicknesses of (7326.7, 7361.6) waves.  The quartz crystal has 34.46 waves retardance while the sapphire crystal has -34.95 waves retardance. The crystals mounted with axes aligned add retardance to become a -0.492 wave retarder.  

This bi-crystalline achromat has e- and o- beam optical path lengths of (12964.6, 12965.1) waves. The internal interface between quartz and sapphire as well as the exit surface reflection will both contribute to the fringes. The fringe period from the combined optical thickness is 0.03nm a wavelength of 400 nm.  There will be additional contributions from the reflections internal to the crystals. As examples, the o- beams contribute fringes with periods of 400 nm / 7360 = 0.054 nm and 400 nm / 5603 = 0.07 nm. In the air-spaced design, there will be an additional contribution coming from multiple internal reflections inside the air gap as well as multiple internal reflections from internal faces to other internal faces.

The Berreman calculus solves the eigenvalue problem which simultaneously satisfies all boundary conditions for the electric field on the entrance and exit of the optic. It is useful for thick crystal retarders to think of the primary contributions to the fringes as the internal reflections with fringe periods scaling as the optical thickness of the crystal.  As an example, Figure \ref{fig:spinor_fringe_period} shows the predicted fringe period for the three different SPINOR modulators using $\lambda^2$/2dn.  We have computed the fast Fourier transforms (FFTs) of various Mueller matrix elements and retardance fits on a wide variety of data sets. For thick crystals, there is excellent agreement between the predicted and calculated fringe spectral period. We will apply this technique in later sections to SPINOR data as well as to many-crystal retarders at other telescopes.

We have created fringe models for the three SPINOR retarders considering several design choices.  Crystals are either optically contacted or air-gapped as well as coated and uncoated. For each model, we can fit a linear retarder Mueller matrix to the Berreman calculus results. For the case of zero incidence angle with exact alignment of the crystals, a linear retarder fit is exact for this optic.  In later sections we discuss non-zero incidence angle and errors in manufacture (crystal orientation, polish) which produces elliptical retardance  (linear + circular retardance). For the simple models of SPINOR bicrystalline achromats, the fit to the linear retardance fast axis orientation is within numerical precision ($<$1e-11) of zero.  The theoretical prediction computed as a series of linear retarders agrees perfectly with the Berreman calculus.

To estimate fringe amplitudes, we apply a simple spectral analysis technique. We have the fringe calculations at a spectral sampling of 1/2,000,000 but the fringes vary on a wavelength scale of a few parts per thousand.  We first subtract the theoretical linear retardance from the fringe model prediction to give linear retardance variation about the design. We then compute the maximum deviation in spectral bins of 50 samples to produce a curve showing the maximum fringe amplitude in narrow bandpasses sampled at 1/40,000.

\begin{figure}[htbp]
\begin{center}
\vspace{-3mm}
\hbox{
\hspace{-1.5em}
\includegraphics[height=11.8cm, angle=0]{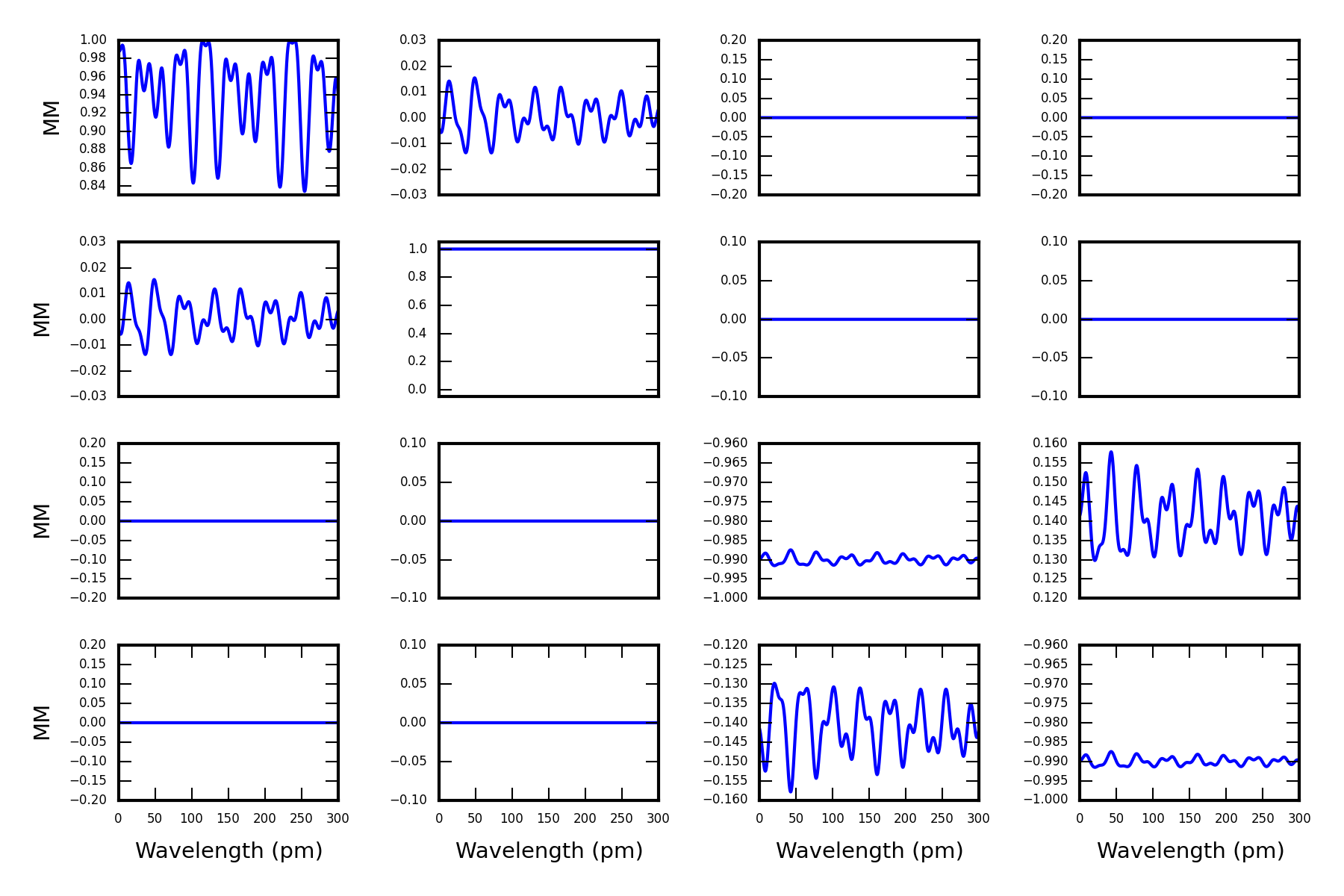}
}
\caption[Mueller Matrix of the SPINOR Modulator] 
{ \label{fig:spinor_modulator_mueller_matrix} 
The Mueller matrix of the SPINOR third-wave bi-crystalline achromatic retarder used as the modulator at a wavelength of 425.7nm. The [0,0] Mueller matrix element has been used to normalize all other matrix elements as in Equation \ref{eqn:MM_IntensNorm}. The retarder has roughly 172$^\circ$ retardance, close to half-wave. Spectral sampling is $\lambda$/$\delta\lambda$ = 2,000,000. This simulation used a single layer quarter-wave thickness of isotropic MgF$_2$ coating to reduce reflections at 550 nm central wavelength. The crystals were air-spaced with a 20 $\mu$m gap.}
\vspace{-8mm}
\end{center}
\end{figure}

Figure \ref{fig:spinor_modulator_linear_retardance_fringe} shows examples of these fringe amplitude estimates for the SPINOR 3/8 wave modulator.  The green curve on the bottom shows an optically contacted, uncoated optic.  The linear retardance varies by roughly $\pm$0.5$^\circ$ to $\pm$1.4$^\circ$ across the 400 nm to 1200 nm wavelength range.  The blue curve shows the same bi-crystalline achromat but with a single layer of isotropic MgF$_2$ as an anti-reflection coating with a central wavelength of 550nm on every surface and a 20$\mu$ air gap between crystals.  The variation in linear retardance for this coated, gapped model is now less than $\pm$0.5$^\circ$ near the AR coated wavelengths but the air gap causes amplitudes to rise over $\pm$1.5$^\circ$ to $\pm$3.0$^\circ$ at long wavelengths.   The Mueller matrix of the air-gapped optic now combines several fringe periods with mixing between the relatively large period air-gap fringe and relatively small period crystal fringes. Figure \ref{fig:spinor_modulator_mueller_matrix} shows the Mueller of the air-gapped, AR-coated SPINOR 3/8 wave modulator in a narrow spectral bandpass from 425.7nm to 426.0nm.  As seen later, this is a bandpass where we have SPINOR observation and is representative of the complexity in analyzing the diattenuation and retardance variation caused by these retarders. The diattenuation and linear retardance are predicted to contain a mix of several spectral periods with amplitudes that depend strongly on the anti-reflection coating performance and the air-gap.

\clearpage

\subsection{SPINOR Observations and Fringe Properties}

We collected observations of sunspots in March of 2016 at the DST using several configurations of the SPINOR spectrograph. We collected a minimum of several hours observing at wavelengths of 426 nm, 431 nm, 458 nm, 514 nm, 526 nm and 614 nm to test six separate wavelengths.

\begin{figure}[htbp]
\begin{center}
\vspace{-3mm}
\includegraphics[width=0.99\linewidth, angle=0]{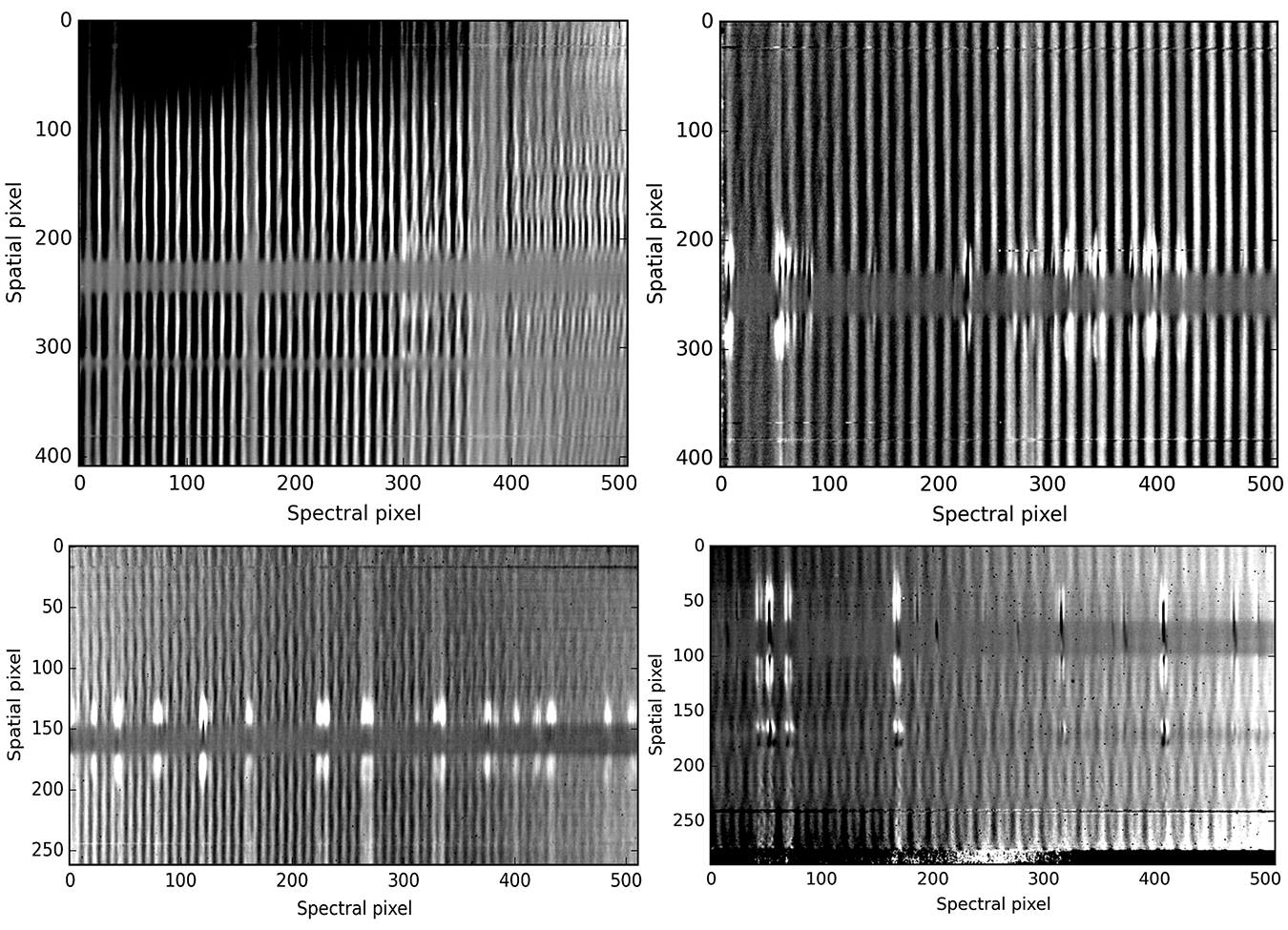}
\caption{Examples of reduced and demodulated Stokes $Q$ measured by SPINOR at the Dunn Solar Telescope while observing a sunspot. The 3/8-wave modulator as well as the 1/2-wave retarder on the polarizing beamsplitter were in the optical path during observations.  In each image, there are grey regions corresponding to the center of the sunspot (horizontal, spectral direction).  There are real Stokes $Q$ signals in several spectral lines seen as spatially localized, saturated white / black regions.  Each image is otherwise covered with large amplitude, quasi-periodic polarization fringes caused by the bi-crystalline retarder. The wavelengths are 425.8nm upper left, 526.2 upper right, 514.2nm lower left, 614.9nm lower right.  The lower two images use a different sensor than the others and have a 256 by 512 pixel format.  The shorter-wavelengths are observed with a 512 by 512 pixel format camera. }
\label{fig:SPINOR_fringe_data}
\vspace{-4mm}
\end{center}
\end{figure}

The 3/8 wave retarder used as a rotating modulator was placed in the converging f/36 beam located just ahead of the SPINOR spectrograph slit.  The data was reduced and processed using the standard data analysis software available at the National Solar Observatory website \footnote{http://nsosp.nso.edu/dst-pipelines}.  Data reduction was performed by Christian Beck.  Example demodulated observations of Stokes $Q$ are shown in Figure \ref{fig:SPINOR_fringe_data} for all six wavelengths.  In each panel of Figure \ref{fig:SPINOR_fringe_data}, there is a sunspot and visible changes in the polarized spectra. Often in the sunspot the intensity drops substantially and the fringe character changes.  There are clear spectral features corresponding to real solar magnetic signatures, but fringes cover and often dominate the entire $Q$ image for most wavelengths.

We analyze the fringes by masking out regions of the image corresponding to the sunspot and real signals.  The SPINOR data was then Fourier transformed to create a power spectrum for the 6 wavelengths observed.

\begin{wrapfigure}{r}{0.65\textwidth}
\centering
\vspace{-3mm}
\begin{tabular}{c} 
\hbox{
\hspace{-1.0em}
\includegraphics[height=7.3cm, angle=0]{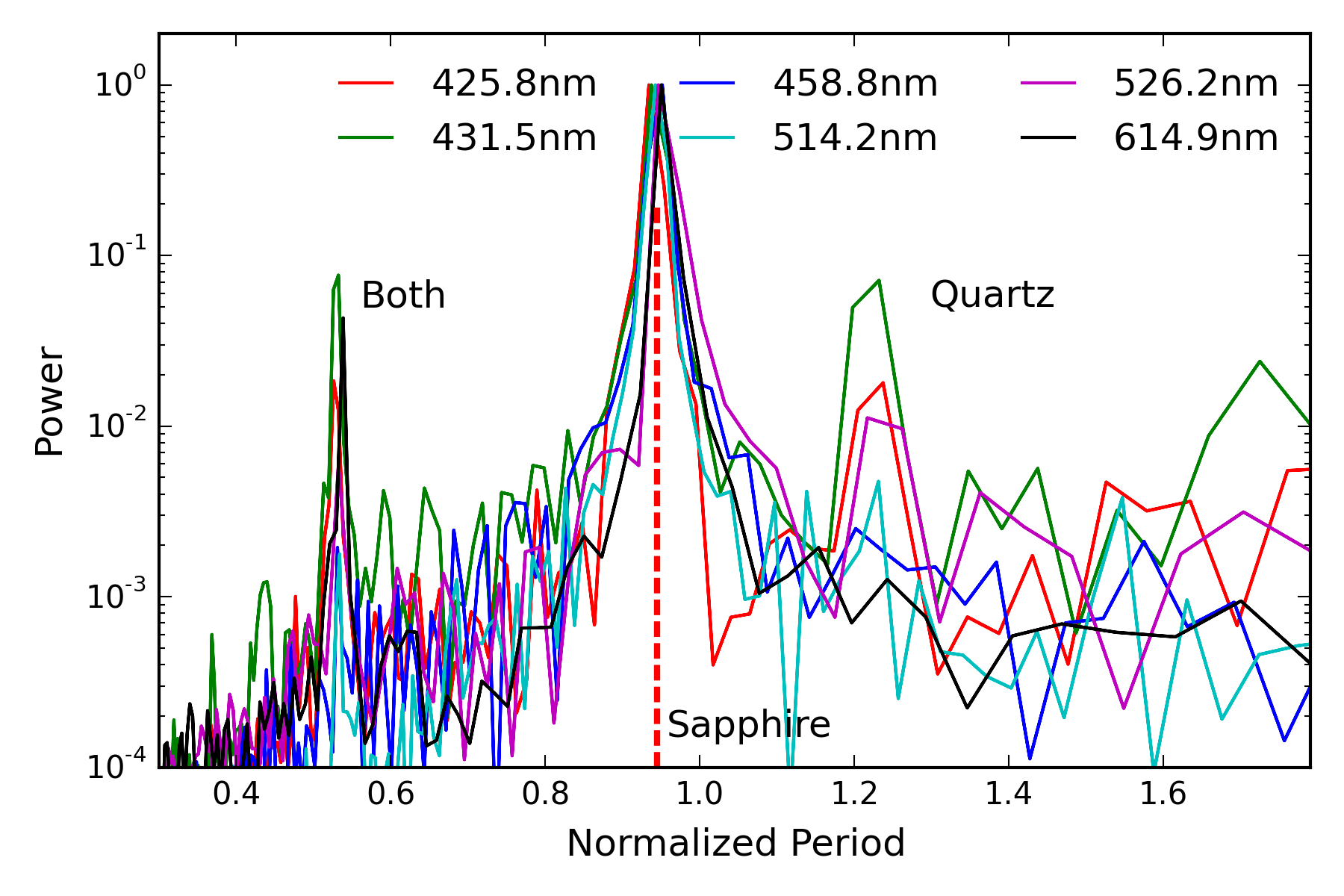}
}
\end{tabular}
\caption[Power in SPINOR Data] 
{ \label{fig:spinor_fft_power} 
The spectral power of the measured Stokes $Q$ fringes in regions without noticeable signals. All six wavelengths are plotted with a different color: 425.8 nm, 431.5 nm, 458.8 nm, 514.2 nm, 526.2 nm and 614.9 nm. Each spectral fringe power spectra had the spectral period normalized by the expected fringe period based on the thickness of the sapphire crystal e- beam ($\lambda^2$ / 2dn). The dominant fringe power is around 94.5\% of the predicted period for the sapphire crystal as seen by the vertical red dashed line. As the quartz piece is thinner and lower index, the fringe caused by the quartz crystal is 1.30 times larger than for the sapphire crystal. The combined crystals (quartz+sapphire) create fringes with periods roughly 0.565 times smaller than the sapphire crystal alone. See text and Table \ref{table:SPINOR_Retarders_DST} for details.}
\vspace{-3mm}
 \end{wrapfigure}

At each wavelength, the effective optical thickness of the retarder is different.  Table \ref{table:SPINOR_Retarders_DST} shows the optical thickness for the 3/8 wave retarder (modulator). We adopt the 3/8 wave modulator crystal thickness scaled from the nominal design values specified by the DST staff.  The numerical solution for the exact retardance specification using our refractive index formulas solved for a crystal quartz thickness of 1.438883026mm and a sapphire crystal thickness of 1.648256206mm. Thickness must be specified to sub-nanometer precision or solved exactly in the Berreman calculus using refractive index formulas as a few microns of crystal typically corresponds to a quarter-wave of retardance. The quartz crystal has a net retardance (bias) of about 21.2 waves at 615 nm while the sapphire crystal has roughly -21.7 waves retardance bias for the SPINOR modulator. 

The first column of Table \ref{table:SPINOR_Retarders_DST} shows the six wavelengths we observed at the DST.  The second column shows optical path for the quartz crystal extraordinary beam (dn/$\lambda$).  The third column shows the sapphire crystal extraordinary beam optical thickness.  The fourth column shows the summed part thickness for the bicrystalline achromat (ignoring the air gap of a few waves thickness). The next two columns show the net retardance for the quartz and sapphire crystals.  The predicted fringe period in picometers ($\lambda^2$/2dn) for the sapphire crystal e- beam is then shown to reference Figure \ref{fig:spinor_fft_power}.  The final column shows the single pixel wavelength sampling for the SPINOR sensor as observed during our campaign.  At short wavelengths, we sample a 31pm fringe period with 3pm sampling giving 10 pixels per fringe. At long wavelengths, we sample at 65pm fringe at 4pm giving 16 pixels per fringe.

To show all power spectra on the same spectral period array, we normalized the x axis by the individual periods predicted for the sapphire crystal fringe at each wavelength.  Both crystals in the optic are a few thousand waves thick with a birefringence of 20 to 30 waves. There is little difference between ordinary and extraordinary beam fringe periods.

\begin{wraptable}{l}{0.65\textwidth}
\vspace{-3mm}
\caption{SPINOR 3/8 Wave Retarder Optical Properties}
\label{table:SPINOR_Retarders_DST}
\centering
\begin{tabular}{l l l l l l l l l}
\hline\hline
					& SiO$_2$	& Al$_2$O$_3$	& Both			& $\phi_{qtz}$	& $\phi_{saph}$& Al$_2$O$_3$& SP	  	\\
$\lambda$	  			& path	  	& path		& path			& Bias		& Bias		& Period 		& samp		\\
nm					& waves		& waves		& waves			& waves		& waves		& pm			& pm			\\
\hline
\hline
425.8	 	 		& 5286.0		& 6867.5		& 12153.5			& 32.03		& -32.51		& 30.86		& 2.78	\\
431.5	 	 		& 5214.2		& 6773.8		& 11988.0			& 31.54		& -32.01		& 31.70 		& 2.68	\\
458.8	 	 		& 4895.9		& 6358.9		& 11254.8			& 29.40		& -29.85		& 35.91		& 3.21	\\
514.2	 	 		& 4357.3		& 5657.2		& 10014.5			& 25.86		& -26.29		& 45.24		& 5.34	\\
526.2	 	 		& 4256.0		& 5525.3		& 9781.4			& 25.21		& -25.63		& 47.40		& 3.16	\\
614.9	 	 		& 3632.5		& 4714.2		& 8347.1			& 21.27		& -21.65		& 64.92		& 4.12	\\
\hline
\hline
\end{tabular}
\vspace{-5mm}
\end{wraptable}

These six fringe spectral periods are listed in Table \ref{table:SPINOR_Retarders_DST}.  As the quartz piece is thinner  (1.44 mm vs 1.65mm) and the refractive index is lower (1.56 vs 1.78), the fringe caused by the quartz crystal is 1.30 times larger than for the sapphire crystal.  Figure \ref{fig:spinor_fft_power} shows the fringe power clearly in several measurements.  The combined crystals (quartz+sapphire) have periods roughly 0.565 times smaller than the sapphire crystal alone.  As an example from Table \ref{table:SPINOR_Retarders_DST}, the quartz crystal optical thickness is $\sim$5290 waves at 425.8 nm while the sapphire crystal is $\sim$6870 waves.  Both crystals together account for $\sim$12,150 waves of optical path.  The fringe period would be 31 picometers for the sapphire crystal, roughly 40 pm for the quartz crystal and 15 pm for both crystals combined.  At this wavelength, SPINOR samples the spectrum with 2.78 pm per pixel.  SPINOR  resolves the shortest period fringe with roughly 5.5 samples per fringe period. The dominant power caused by the sapphire crystal is sampled at 11.1 pixels per fringe period.

\section{DKIST Optics: Retarders, Windows \& Dichroics}

The DKIST optical path has several windows, beamsplitters (BS) and crystal-stack retarders with various coatings. In a recent paper, we outlined the beam path with a focus on the reflective optics \cite{Harrington:2017dj}.  The telescope feeds an f/53 beam to a coud\'{e} laboratory which is then split between several systems. There is a BS that feeds the adaptive optics wavefront sensor. This BS substrate is 43mm thickness of Infrasil, uncoated on the front side and broad-band anti-reflection coated (BBAR) on the back side. We call this part WFS-BS1 as it is the first BS in the optical path to feed the wavefront sensor. In addition to this BS, there are several dichroic BSs. These dichroics are the Facility Instrument Distribution Optics (FIDO) that have more complex coatings for reflecting and transmitting different wavelengths. These dichroic substrate windows also produce fringes with a more varied wavelength dependence of the amplitude due to the coatings. 

\begin{wrapfigure}{r}{0.65\textwidth}
\centering
\vspace{-4mm}
\begin{tabular}{c} 
\hbox{
\hspace{-1.5em}
\includegraphics[height=7.1cm, angle=0]{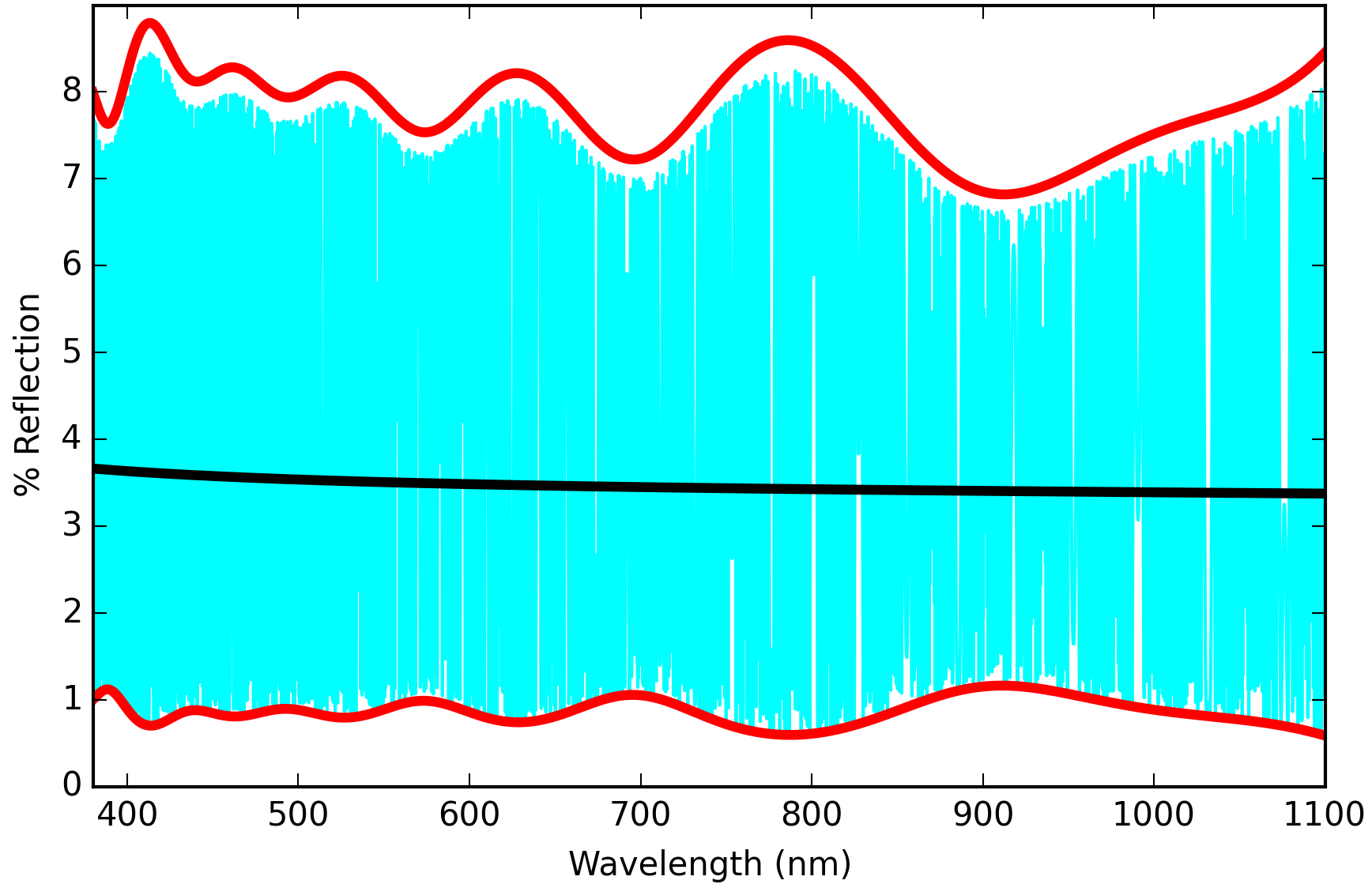}
}
\end{tabular}
\caption[Fringes] 
{ \label{fig:reflectivity_wfsbs1} 
The transmission for WFS-BS1. The Berreman calculus was used to predict fringes for our 43mm thick Infrasil BS.  The front surface was uncoated giving a reflection of $\sim$3.5\%.  The back surface was coated with an 18-layer formula from Infinite Optics that gives $\sim$1\% reflection across the 380nm to 1100nm wavelength range.  The cyan curve shows Berreman fringe predictions for transmission at a spectral sampling of $\delta \lambda$ / $\lambda$ of 5,000. Fringes are unresolved but oscillate between 8\% and 1\%.  The black curve shows the Berreman-predicted transmission of the uncoated air to Infrasil interface. Red curves show the simple analytic formula of $I_{tot} = (\sqrt{I_{front}}  \pm  \sqrt{I_{back}})^2$.}
\vspace{-3mm}
 \end{wrapfigure}

To show how anti-reflection coatings influence fringes on a simple BS, we focus on WFS-BS1.  The substrate for this optic is specified to have an intrinsic strain  (stress birefringence) of less than 5nm per cm of optical path using Heraeus Infrasil 301 as our nominal material choice. With this low stress value, we model the part as an isotropic glass with no polarization influence.  The optic is placed at an incidence angle of 15$^\circ$ to the incoming beam. With this incidence angle, the Fresnel reflection off the front and back surfaces does introduce linear polarization. As we showed above, even a simple tilted glass substrate can introduce polarization fringes in transmittance, diattenuation (polarizance) and also retardance because of the coherent interference between the S- and P- beams from both front and back surfaces. 

To compute the likely fringe amplitude, the expected electric field amplitudes must be added or subtracted coherently.  As a first-order approximation for highly transmissive optics, we can simply consider the reflection from the back surface interfering coherently with the front surface.  The initial reflection has an intensity of 3.5\% at the air to Infrasil interface.  The transmitted beam reflects off the back surface and then transmits again through the front surface with an intensity that scales with the back surface reflectivity twice multiplied by the air to Infrasil transmission.  This is roughly 96.5\% x 1.0\% x 96.5\% $\sim$ 0.93\%.

With this assumption of a singly-reflected beam causing interference, we can predict the total intensity for constructive and destructive interference at each wavelength. We can compute the interference using the predicted reflection and transmission coefficients as $I_{tot} = (\sqrt{I_{front}}  \pm  \sqrt{I_{back}})^2$ giving fringes as 4$(\sqrt{I_{front}} \sqrt{I_{back}})$.  For our nominal 3.5\% front surface reflection interfering with a 0.9\% back surface contribution, we see field amplitudes of 0.187 $\pm$ 0.097 which gives predicted intensities of roughly 8\% to 1\% for the transmission fringes.  

Figure \ref{fig:reflectivity_wfsbs1} compares the Berreman calculus prediction for transmission to the simple analytical equation presented in the previous paragraph. The cyan curve shows unresolved transmission fringes that rapidly oscillate between 1\% and 8\%.  We use our simple equation for interference from a single reflection off the back surface to compute the red curves.  Note that the Berreman calculus is an eigenvalue method that automatically accounts for the infinite number of internal reflections.  As such, our single-reflection estimate is slightly in error from the amplitudes predicted by the Berreman calculus.

\begin{figure}[htbp]
\begin{center}
\vspace{-1mm}
\includegraphics[width=0.99\linewidth, angle=0]{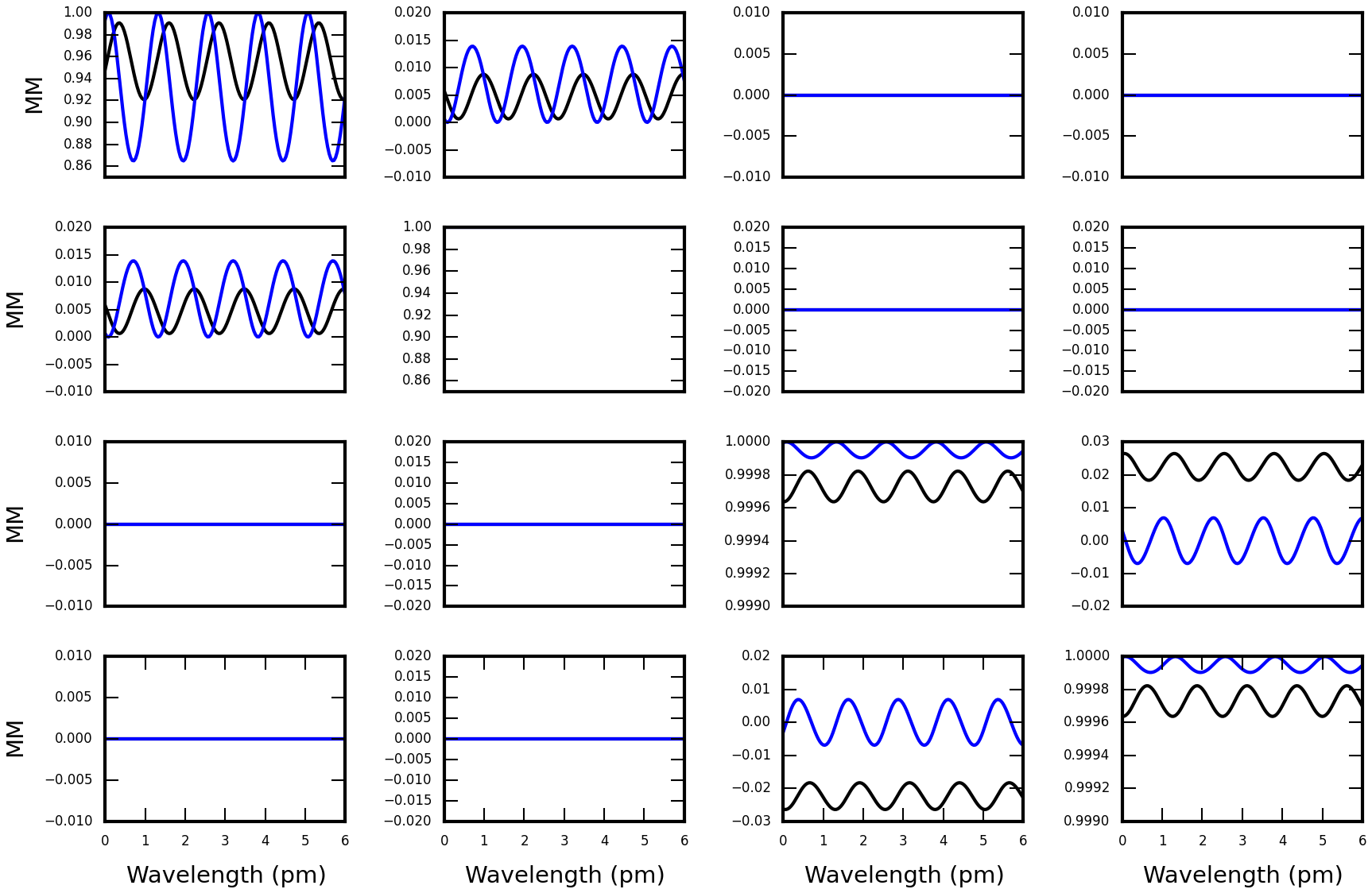}
\caption{The Mueller matrix of WFS-BS1 sampled at  $\delta \lambda$ / $\lambda$ of 1,000,000 using the Berreman calculus. The [0,0] Mueller matrix element has been used to normalize all subsequent elements to separate transmission effects from polarization as in Equation \ref{eqn:MM_IntensNorm}. The [1,1] element $QQ/II$ is identically 1 in this plot. There are transmission, diattenuation (polarizance) and retardance fringes in this BS. The blue curve shows the uncoated 43mm thick Infrasil beamsplitter with roughly 3.5\% reflection on both entrance and exit surfaces.  The black curve shows the Berreman model with an uncoated front surface and an 18-layer broad-band anti-reflection coating from Infinite Optics applied to the back surface.  The nominal wavelength is 400nm with a fringe period of roughly 1.3pm.  The anti-reflection coating reduces the amplitude of all fringes: transmittance, diattenuation (polarizance) and retardance. The $UV$ and $VU$ terms show a net change in retardance as expected for adding a coating on a tilted optic though retardance fringe amplitude is slightly reduced.}
\label{fig:MM_wfsbs1}
\vspace{-4mm}
\end{center}
\end{figure}

This seemingly simple BS also shows fringes in diattenuation and retardance.  Figure \ref{fig:MM_wfsbs1} shows the Mueller matrix for the optic sampled at  $\delta \lambda$ / $\lambda$ of 1,000,000.  As expected, the BS shows polarization fringes with a period consistent with the optical thickness.  Fused silica has a refractive index of 1.47 at 400nm wavelength.  A 43mm thick substrate has an effective thickness of 158025 waves.  This gives an expected fringe spectral period of 1.3pm.  This matches the fringes seen in Figure \ref{fig:MM_wfsbs1} quite well.

The fringe amplitudes for transmission, diattenuation and retardance all decrease by a factor of roughly 2x when the 1\% anti-reflection coating is used. The lower right hand 2x2 matrix of Figure \ref{fig:MM_wfsbs1} shows that the retardance change involves a fringe amplitude reduction and also a change in the part retardance, as seen by the offset of the $UV$ and $VU$ terms. If we have an uncoated front surface and a 1\% reflectivity coated back surface, we get fringes of 4$(\sqrt{0.035} \sqrt{0.01})$ $\sim$7.5\%.  Even if we put a 1\% coating on both surfaces, the intensity fringe would be 4$(\sqrt{0.01} \sqrt{0.01})$ $\sim$4.0\%.  With coatings of 0.7\% reflectivity, fringes are still 4$(\sqrt{0.007} \sqrt{0.007})$ $\sim$2.8\% amplitude. These fringes are subject to several thermal instabilities.  The physical expansion of the optic through the Coefficient of Thermal Expansion, the refractive index temperature dependence through the thermo-optic coefficient, and the birefringence change with temperature all play a role in the fringe dependence on temperature.

\subsection{DKIST Calibration Retarder}

\begin{wraptable}{l}{0.30\textwidth}
\vspace{-3mm}
\caption{ViSP SAR Retarder}
\label{table:ViSP_SAR_Design}
\centering
\begin{tabular}{l l l}
\hline
\hline
Material	& Thickness	& $\theta$		\\
		& $\mu$m		&  deg.		\\
\hline
\hline
AR 	& 0.0944		& -	\\
FS	& 10000.0		& -	\\
AR 	& 0.0944		& -	\\
Oil	& 10.0		& -	\\
AR 	& 0.0944		& -	\\
Qtz	& 2122.1		& 0	\\
AR 	& 0.0944		& -	\\
Oil	& 10.0		& -	\\
AR 	& 0.0944		& -	\\
Qtz	& 2099.1		& 90	\\
AR 	& 0.0944		& -	\\
Oil	& 10.0		& -	\\
AR 	& 0.0944		& -	\\
Qtz	& 2132.4		& 70.25	\\
AR 	& 0.0944		& -	\\
Oil	& 10.0		& -	\\
AR 	& 0.0944		& -	\\
Qtz	& 2099.1		& 160.25	\\
AR 	& 0.0944		& -	\\
Oil	& 10.0		& -	\\
AR 	& 0.0944		& -	\\
Qtz	& 2122.1		& 0	\\
AR 	& 0.0944		& -	\\
Oil	& 10.0		& -	\\
AR 	& 0.0944		& -	\\
Qtz	& 2099.1		& 90	\\
AR 	& 0.0944		& -	\\
Oil	& 10.0		& -	\\
AR 	& 0.0944		& -	\\
FS	& 10000.0		& -	\\
AR 	& 0.0944		& -	\\
\hline
\hline
\end{tabular}
See text for details.
\vspace{-7mm}
\end{wraptable}

In this section we show models for the three DKIST calibration retarders designed to be 0.25 wave linear retarders in specific bandpasses. Each calibration retarder contains 6 individual retarder crystal plates.  This increases the complexity of the retarder fringe model and shows the interference effects from multiple internal interfaces in birefringent media.  DKIST will produce three additional modulating retarders which are designed to be elliptical.  We will focus on the three linear retarders for simplicity.

A common retarder design tool was introduced by Pancharatnam \cite{Pancharatnam:1955iw} to make a super-achromatic retarder as a combination of three retarders. By using 3 retarders together, the wavelength range for achromatic linear retardance is greatly increased. For our particular application, we make each of these three retarders as subtraction-pairs of crystal plates.  This way there are six total crystals, following the Pancharatnam design where each of the three linear retarders are now comprised of two crystals used in subtraction.

There are many degrees of freedom if one chooses different materials, retardance values and orientations for all six crystals.  The Pancharatnam designs can be simplified by choosing just one or two materials and also by making the outer two retarders identical. This simple design uses an A-B-A type alignment where the two outer crystal pairs are mounted with their fast axes aligned. Provided the bi-crystalline pairs are treated as perfect linear retarders, there is a simple theoretical formula for the linear retardance of such an A-B-A design. 

If we take the retardance of the A crystals as $\delta_A$ and the B crystals as $\delta_B$, and the relative orientation between the A and B crystal pairs as $\theta$, we can write the formula for the resulting superachromatic optic retardance ($\Delta$) and fast axis orientation ($\Theta$) as in Equations \ref{pan_stack1} and \ref{pan_stack2} \cite{Pancharatnam:1955iw}.

Often, a further constraint is to make all three pairs identical.  There is still an orientation offset between the inner B pair and the outer A pairs. This way, a simple Pancharatnam design would only use two materials (such as Quartz and MgF$_2$ crystal) and a manufacturer would only polish each material to one specific thickness. This way, the retarder has three identical subtraction pairs achromats with an orientation of [0$^\circ$,  X$^\circ$, 0$^\circ$] and only two thicknesses to vary for a 3-dimensional optimization problem.  Each DKIST calibration retarder was designed using six crystals grouped into three pairs using either all quartz or all MgF$_2$ crystals.  Each 2-crystal group is a subtraction pair. 

\begin{wrapfigure}{r}{0.45\textwidth}
\centering
\vspace{-12mm}
\begin{equation}
\label{pan_stack1}
\cos \frac{\Delta}{2}  =  \cos \frac{\delta_B}{2}  \cos \delta_A  - \sin \frac{\delta_B}{2}  \sin \delta_A  \cos 2\theta  
\end{equation}
\begin{equation}
\cot 2\Theta =  \frac{ \sin \delta_A  \cot \frac{\delta_B}{2}  +  \cos \delta_A \cos  2\theta }   {  \sin 2\theta }
\label{pan_stack2}
\end{equation}
\vspace{-10mm}
\end{wrapfigure}

\begin{figure}[htbp]
\begin{center}
\vspace{-3mm}
\hbox{
\hspace{-1.5em}
\includegraphics[height=11.8cm, angle=0]{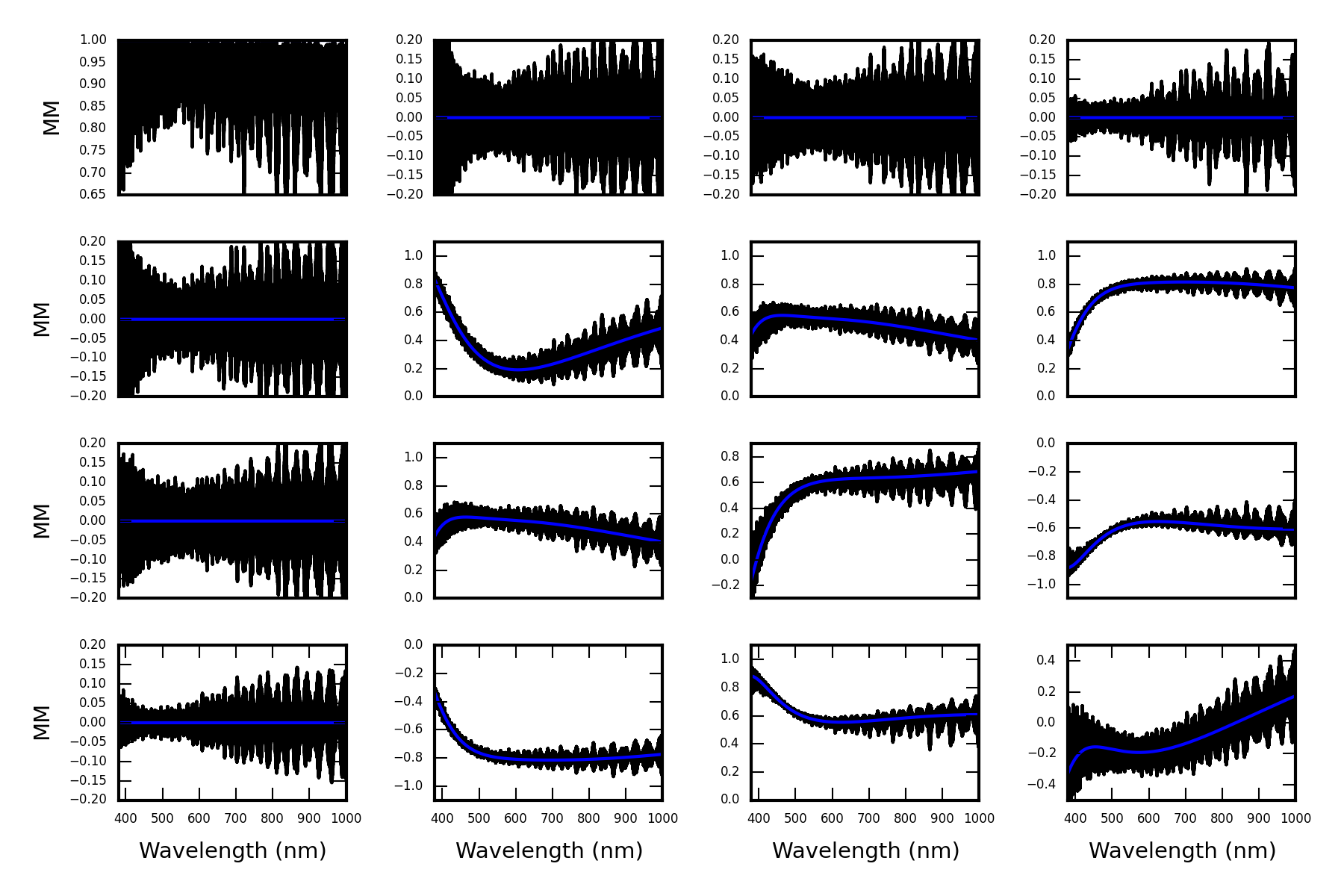}
}
\caption{The Mueller matrix for the visible-wavelength quartz calibration retarder including the two 10mm thick windows as listed in Table \ref{table:ViSP_SAR_Design}. The [0,0] element is used to normalize all subsequent elements to separate transmission effects from polarization as in Equation \ref{eqn:MM_IntensNorm}. Black shows the Mueller matrix computed by the Berreman calculus.  Blue shows the theoretical Mueller matrix computed from the three ideal linear retarders in the A-B-A design. No  fringes are included in the blue curve.  The blue curve follows the $QUV$ to $QUV$ elements as expected.  The Berreman code matches the simple {\it A-B-A stack of linear retarders} theory but with fringes in the first row and column as well as significant retardance fringes in all $QUV$ to $QUV$ elements. }
\label{fig:ViSP_SAR_MM_VIS}
\vspace{-6mm}
\end{center}
\end{figure}

Crystals were mounted with their fast and slow axes crossed in order to perform the subtraction. The quartz plates are all designed to be 30 waves of retardance thick while the MgF$_2$ crystals are 40 waves retardance.  Both the quartz and MgF$_2$ crystal thicknesses end up being a little over 2 mm physical thickness.  As the Berreman calculus includes all interference from all interfaces, we show here the impact of adding anti-reflection coatings, air gaps, refractive index matching oils, optical contacting and / or the inclusion of cover windows often used to achieve specific flatness and polish specifications. Figure \ref{fig:ViSP_SAR_MM_VIS} shows the Berreman-derived Mueller matrix including all optics, coatings and oil from Table \ref{table:ViSP_SAR_Design}

For our large retarders, the 120 mm diameter and 2 mm thickness prevented optical contact as a viable method for bonding the crystals together. Our calibration retarders must work in a 300W beam of 105 mm diameter with substantial flux at short and long wavelengths. This heat load, UV flux and high aspect ratio places stringent requirements on any materials including coatings and oils placed between the crystals.  

DKIST performed tests on several types of refractive-index matching oil to ensure survivablilty and performance when observing wavelengths from 380 nm to 5000 nm while being exposed to non-zero flux for wavelengths as short as 300 nm.  Other epoxies and cements were also considered and ruled out as they would risk catastrophic failure of the part, should the material every degrade over the 40-year lifetime of the observatory.

There were stringent restrictions on transmitted wavefront error and also beam deflection for these optics. Beam deflection in particular causes image motion as the optic rotates. For rotating retarders used as modulators, this is a major source of error in processing polarimetric image sequences. In response to all these concerns, the DKIST team decided to require relatively thick (10 mm) cover windows on both exit and entrance interfaces to the crystal stacks. These windows are thought to assist with wavefront error and beam deflection, but we show here there are also strong consequences for polarization fringe amplitudes and spectral periods. There are several other strong impacts to beam wobble and heating from including cover windows. 

The calibration retarders were designed as super-achromatic retarders (SARs) with a traditional quarter-wave linear retarder design goal. Of course, calibration does not require only linear retardance and does not require particularly chromatic behavior \cite{1990JOSAA...7..693G,Kupinski:2014ek,2012OptL...37.1097L, LaCasse:2011kv, Twietmeyer:2005bz, delToroIniesta:2000cg, Tomczyk:2010wta,Wijn:2011wt,deWijn:2010fh,2010SPIE.7735E..4AD, Compain:1999do, Harrington:2010km}. Elliptical, chromatic retarders are suitable, but the DKIST project decided to follow this common quarter-wave linear strategy at the time procurement decisions were made. 

There are three subtraction crystal pairs designed to work in the combination A-B-A when used as a calibration (linear) retarder.  This design strategy and the crystal thicknesses were also influenced by other manufacturing reasons.  The elliptical retarders used as a rotating modulator in each DKIST instrument covering the same wavelength range could be made using B-A-B crystals with a change of orientation.  In this way, we only need to manufacture crystals of two different thicknesses (A, B) but can make linear retarders and elliptical (efficient) rotating modulators using relatively few design parameters.

\subsection{Quartz Visible Wavelength Retarder}

The first retarder we consider for DKIST is nominally designed to cover 380 nm to 1100 nm wavelength range.  The Visible Spectro-Polarimeter (ViSP) instrument spans roughly this wavelength range. The DKIST project often refers to this retarder as the ViSP SAR. However, this optic is useful over a wide wavelength range and can be used by any instrument. It will be used by other instruments in the same wavelength range such as the Visible Tunable Filter (VTF) or the Diffraction-Limited Near-Infrared Spectropolarimeter (DL-NIRSP).  As an example, all DKIST instruments should be using the 854 nm bandpass for some activities and all instruments can use this retarder for calibration at this wavelength.

\begin{figure}[htbp]
\begin{center}
\vspace{-3mm}
\includegraphics[width=0.99\linewidth, angle=0]{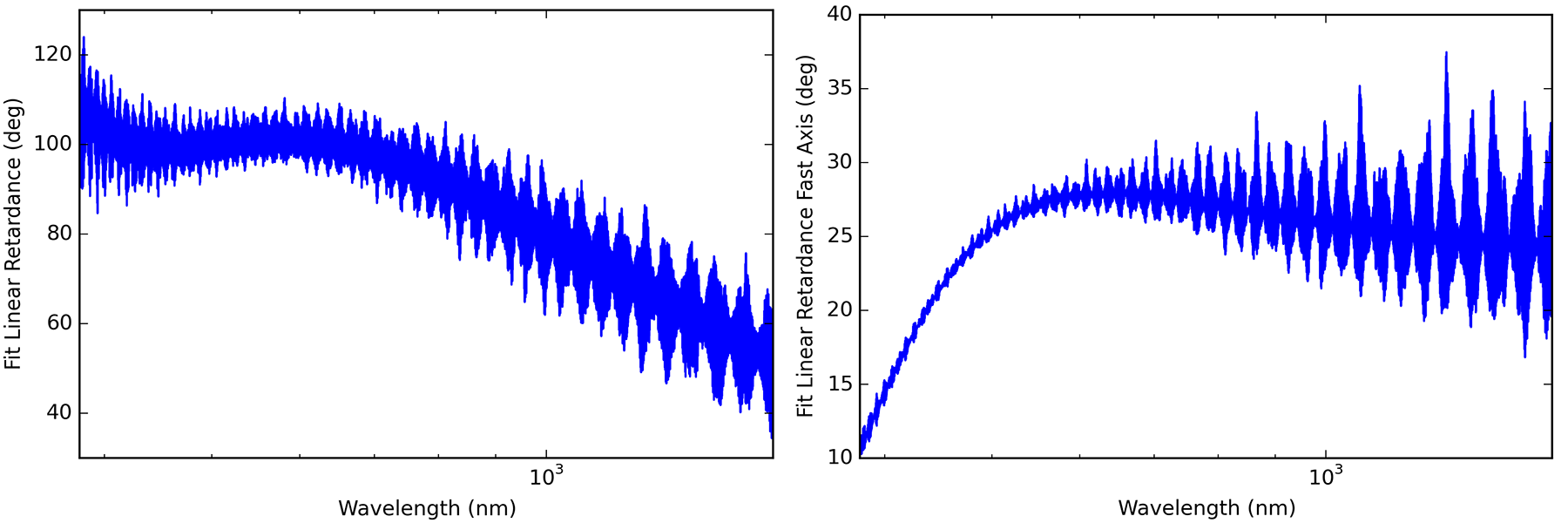}
\caption{The best fit linear retardance and fast axis to the Berreman calculus model of the visible-wavelength quartz retarder design with all optical elements listed in Table \ref{table:ViSP_SAR_Design}. The wavelength range goes from 380nm to 1600nm. The linear retardance amplitude is close to a quarter-wave (90$^\circ$ magnitude) throughout the visible wavelength range and the fast axis orientation changes by about 20$^\circ$ as in the theoretical design. The amplitude of the linear retardance fringes is 10$^\circ$ to 20$^\circ$ in magnitude. Note that the fringes are elliptical and a similar magnitude of circular retardance is not captured in this fit.}
\label{fig:visp_sar_fit_linear}
\vspace{-5mm}
\end{center}
\end{figure}

For this quartz visible-range calibration retarder, the crystal plates were designed to be 30 waves retardance thick with the A- bicrystalline achromat to be polished to 0.476 waves linear retardance at 633.443 nm wavelength.  The B- bicrystalline achromat was designed to be polished to 0.328 waves linear retardance at the same wavelength. By orienting these three bicrystalline achromats as A-B-A with orientations of (0$^\circ$, 70.25$^\circ$, 0$^\circ$) a retarder that is close to a quarter-wave linear retardance is created.  The two cover windows chosen for this optic are 10 mm thickness of Heraeus Infrasil 302. 

When performing a Berreman calculation, we also need to specify all other materials in the stack.  The properties of the ViSP six-crystal superachromatic retarder (SAR) for calibration are shown in Table \ref{table:ViSP_SAR_Design}.  Each of the crystals is about 2 mm thick.  

The refractive index matching oil has been measured on several substrates and is typically in the range of 5 $\mu$m to 15 $\mu$m thick.  We use 10 $\mu$m physical thickness and a constant refractive index of 1.3 for the present modeling.  The Cauchy equation on the data sheet for the oil does show an almost constant n=1.3 across the full wavelength range.  We use a 94.4 nm physical thickness of isotropic MgF$_2$ for the anti-reflection coating to simulate the central wavelength 525 nm specified for the AR coatings.

From the Berreman calculus Mueller matrix model, we can derive the retardance and diattenuation.  Figure \ref{fig:visp_sar_fit_linear} shows fits to the linear retardance components of the Mueller matrix.  The linear retardance follows the predicted design and illustrates how the interference fringes change strongly as the many layers from Table \ref{table:ViSP_SAR_Design} contribute backwards reflected radiation.

As expected for a many-layer part, there is a substantial circular retardance in the fringes.  The residual Mueller matrix values in the $QUV$ to $QUV$ terms are up to 0.05 when comparing best-fit linear retarder models to the fully elliptical Berreman fringe model.  An elliptical retarder fit to this model completely reproduces these terms, showing that the elliptical retardance of the fringe is easily detected and fit.

\subsection{MgF$_2$ Calibration Retarder}

A six-crystal MgF$_2$ retarder was designed as a linear quarter-wave retarder the 2500 nm to 4000 nm wavelength range. The DKIST instrument called CryoNIRSP presently plans to have filters covering at least 1000 nm to 4600 nm. This infrared wavelength range required the use of an IR-transparent material. Obvious choices included sapphire and MgF$_2$ crystals.  Given the manufacturing feasibility some years ago, the DKIST team decided to proceed with a design that included only MgF$_2$ crystals and used CaF$_2$ windows to achieve the required flatness and polish specifications. The two CaF$_2$ cover windows are 10 mm thick each.

\begin{figure}[htbp]
\begin{center}
\vspace{-1mm}
\includegraphics[width=0.99\linewidth, angle=0]{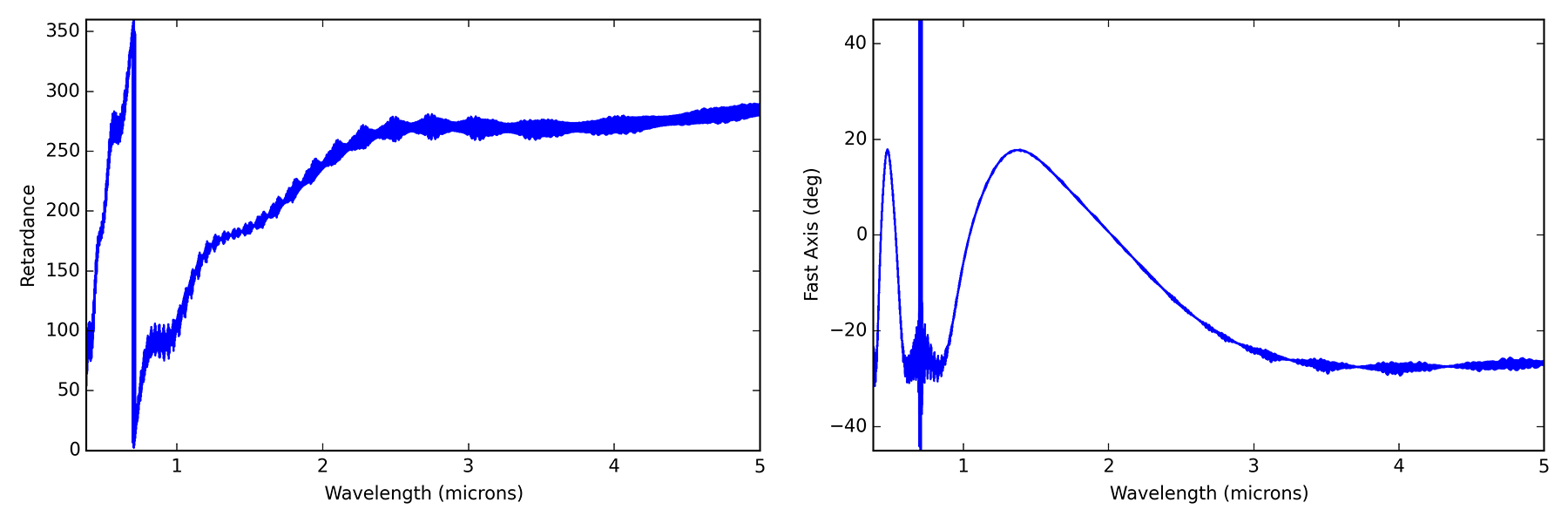}
\caption{The fit linear retardance and fast axis to the CryoNIRSP SAR design which includes 6 crystals, 2 AR-coated cover windows and oil between all interfaces. The left panel shows linear retardance, the right panel shows the fast axis orientation. The design wavelength range was 2500 nm to 4000 nm.  The linear retardance drops through zero and changes sign at short wavelengths.  Near zero, the definition of the fast axis is problematic and noise amplification leads to a vertical line. For clarity, we have wrapped the linear retardance around 360$^\circ$.}
\label{fig:cryo_sar_fit_linear}
\vspace{-2mm}
\end{center}
\end{figure}

This retarder uses the same A-B-A style of three bicrystalline achromats but with MgF$_2$ crystals instead of quartz. The thickness are (2273.14$\mu$m, 2153.11$\mu$m) for A and (2333.21$\mu$m, 2153.11$\mu$m) for B. These correspond to net retardance of 40 waves per plate. The A-B-A pairs are polished to net retardances of 2.230, 3.346, 2.230 waves retardance at 633.443 nm.  The orientations for the bicrystalline pairs are (0$^\circ$, 107.75$^\circ$, 0$^\circ$). We used the revised birefringence measurements from Sueoka as directly measured over a wide wavelength range \cite{Sueoka:2016vo}.

The six crystals combine to a thickness of about 13mm and we use the same 10 $\mu$m thickness for the refractive index matching oil between the crystals.  The anti-reflection coating modeled here used a central wavelength of 1500 nm and we computed a physical thickness of 271.7 nm for the isotropic MgF$_2$ coating applied on all just the 4 window surfaces.  We show in later sections the impact of thicker, thinner and no coating.  The actual design coating has a central wavelength of 2500 nm and a corresponding thickness of 452.9 mm but it is unclear yet if this thick of a coating is feasible.  As the MgF$_2$ crystals have low refractive index and our oil has a close matching refractive index of around 1.3, anti-reflection coatings are not considered for the MgF$_2$ crystals.

We computed the Mueller matrix using the Berreman calculus and fit a linear retarder model.  In contrast to the quartz retarders, this retarder has a strong change in retardance at visible wavelengths as it was optimized for achromatic performance in the 1500 nm to 4000 nm wavelength range.  Figure \ref{fig:cryo_sar_fit_linear} shows the linear retardance fits.  The linear retardance magnitude is near an integer multiple of quarter-wave retardance in the 2500 nm to 4000 nm wavelength range as expected for the relatively thick A-B-A design at (2.230, 3.346, 2.230) waves retardance bias per plate pair.  At shorter wavelengths, the retardance of the optic drops in magnitude to zero near 705nm.  The sign of the retardance changes and the magnitude then continues to change.

\subsection{Design Choices: Trade-Offs for Removing Retarder Windows}

When designing retarders, there are several decisions that can benefit from a quantitative trade off of polarization fringe properties against wavefront error, beam deflection, cost, coating thickness, and other system level choices. 

\begin{figure}[htbp]
\begin{center}
\vspace{-2mm}
\includegraphics[width=0.99\linewidth, angle=0]{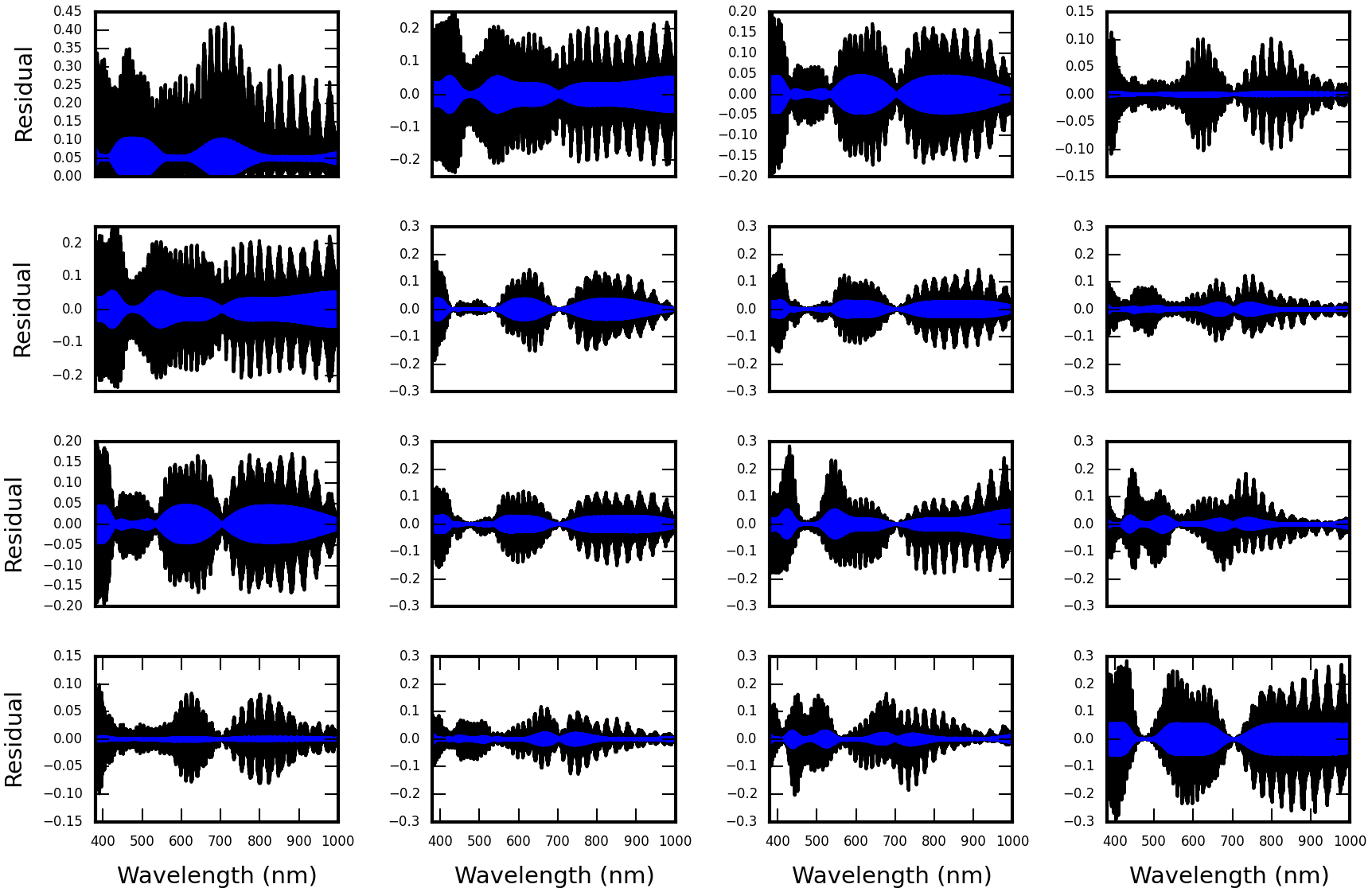}
\caption{The polarization fringes for window-covered (black) and no-window (blue) designs for the MgF$_2$ retarder. The nominal design used CaF$_2$ cover windows (black) on top of the six MgF$_2$ crystal retarders. The CaF$_2$ window to air interfaces are anti-reflection coated with {\it isotropic} MgF$_2$ at a central wavelength of 1500nm (nominal design). When the cover windows are removed (blue), the fringes see much less optical path, the diattenuation amplitudes are substantially reduced in the $I$ to $QUV$ and $QUV$ to $I$ matrix elements. The retardance fringes are also substantially reduced as seen in the $QUV$ to $QUV$ elements. The theoretical Mueller matrix was subtracted from all Berreman-computed elements to isolate the fringes. The [0,0] element ($II$) was used to normalize all elements and separate transmission effects from polarization effects as in Equation \ref{eqn:MM_IntensNorm}. }
\label{fig:cryo_sar_window_cover_fringe_impact}
\vspace{-3mm}
\end{center}
\end{figure}

Often, there are restrictions on the transmitted wavefront quality, polish and beam deflection that impose challenges for relatively high aspect ratio crystals. The DKIST retarders are 2mm thick by 120mm across for a 60:1 aspect ratio. With thin crystal plates, there is difficulty manufacturing a part with low wavefront distortion, retardance uniformity of $<$0.01 waves, environmental insensitivity and benign response to heating through optical absorption. Often times, a solution is to put cover windows on the part. While windows provide many benefits, the impact on polarization fringes must be considered as a trade against potential benefits.

\begin{figure}[htbp]
\begin{center}
\vspace{-1mm}
\includegraphics[width=0.99\linewidth, angle=0]{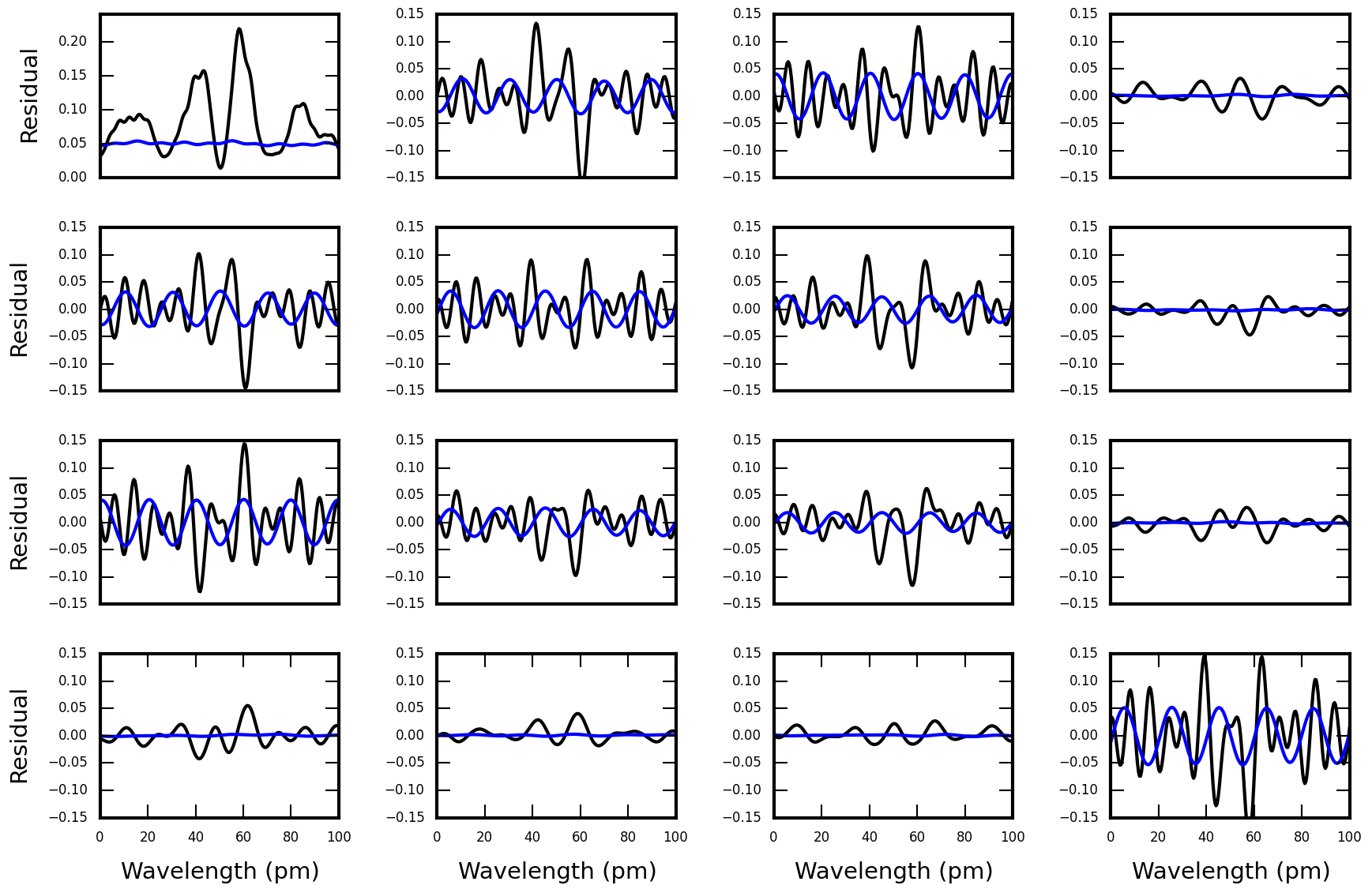}
\caption{The residual Mueller matrix differences between an ideal retarder A-B-A stack Mueller matrix model and the Berreman fringed Mueller matrix model. The [0,0] matrix element $II$ has been used to normalize all subsequent Mueller matrix elements to separate transmission effects from polarization. The nominal window-covered design is shown in black and the no-window uncovered crystal model is shown in blue for the MgF$_2$ calibration retarder.  This Figure shows a narrow wavelength band from 854nm to 854.1nm covering a 100pm bandpass where many DKIST instruments intend to observe. When the cover windows are removed (blue), the fringes see much less optical path and the fringe amplitudes are substantially reduced.}
\label{fig:cryo_sar_window_cover_fringe_impact_zoom_854}
\vspace{-2mm}
\end{center}
\end{figure}

Fringes may be spectrally resolved, spatially variable and can be minimized with coatings. Having fringe predictions to include with other performance metrics allows more realistic systems engineering. For the specific case of the DKIST calibration retarders, the high flux f/13 beam at the Gregorian focus presents special challenges. The set of DKIST polarimetric optics must work with a diffraction limited large aperture optical system in a 300 watt beam with a 105mm clear aperture. The optics must be mounted and rotate near focal planes with substantial flux to wavelengths shorter than 320nm. The set of optics must be used from 380nm to as long as 5000nm with good efficiency and often simultaneous operation of many instruments. This imposes constraints on wavefront error, polishing uniformity, wedge and beam deflection, surface polish and thermal response to name a few.  In this section, we outline a few typical choices involving inclusion or removal of cover windows that may be encountered in design of retarders for large aperture telescopes.

As an example of Berreman fringe predictions, Figure \ref{fig:cryo_sar_window_cover_fringe_impact} shows the computed Mueller matrix for the DKIST MgF$_2$ calibration retarder with and without the 10mm thick CaF$_2$ cover windows. The black curve shows the window-covered optic with transmission fringes ranging up to 0.4.  Diattenuation terms have amplitudes over 0.2 and the $QUV$ to $QUV$ retardance elements of the Mueller matrix are fringing over 0.2.  The blue curve shows the part without these coated CaF$_2$ cover windows. The transmittance, diattenuation and retardance fringes all decrease by factors of 2 to $>$3 when removing the CaF$_2$ cover windows with intensity fringes always at or below 10\%.  

The spectral frequency and complexity of the fringe pattern also increases strongly when cover windows are added. The increase in optical path from cover windows also increases the sensitivity of the part to thermal issues.  Figure \ref{fig:cryo_sar_window_cover_fringe_impact_zoom_854} shows the DKIST calibration retarder with and without the 10mm thick CaF$_2$ cover windows in a narrow wavelength region starting at 854nm covering 0.1nm. In Figure \ref{fig:cryo_sar_window_cover_fringe_impact_zoom_854} the blue curve again shows the optic with cover windows.  Intensity fringes are seen in the $II$ matrix element with amplitudes of 0.2. The blue curve shows the part without CaF$_2$ cover windows. The transmission fringes are nearly zero, the diattenuation decreases amplitude by over 2x in $IQ$ and $QI$.  The circular polarizance $VI$ and $IV$ go to nearly zero without windows.  The complexity and spectral content of the fringes greatly decreases without windows.  These amplitude and frequency decreases are all strong motivations to not use cover windows.

\begin{wrapfigure}{l}{0.60\textwidth}
\centering
\vspace{-5mm}
\begin{tabular}{c} 
\hbox{
\hspace{-1.2em}
\includegraphics[height=6.9cm, angle=0]{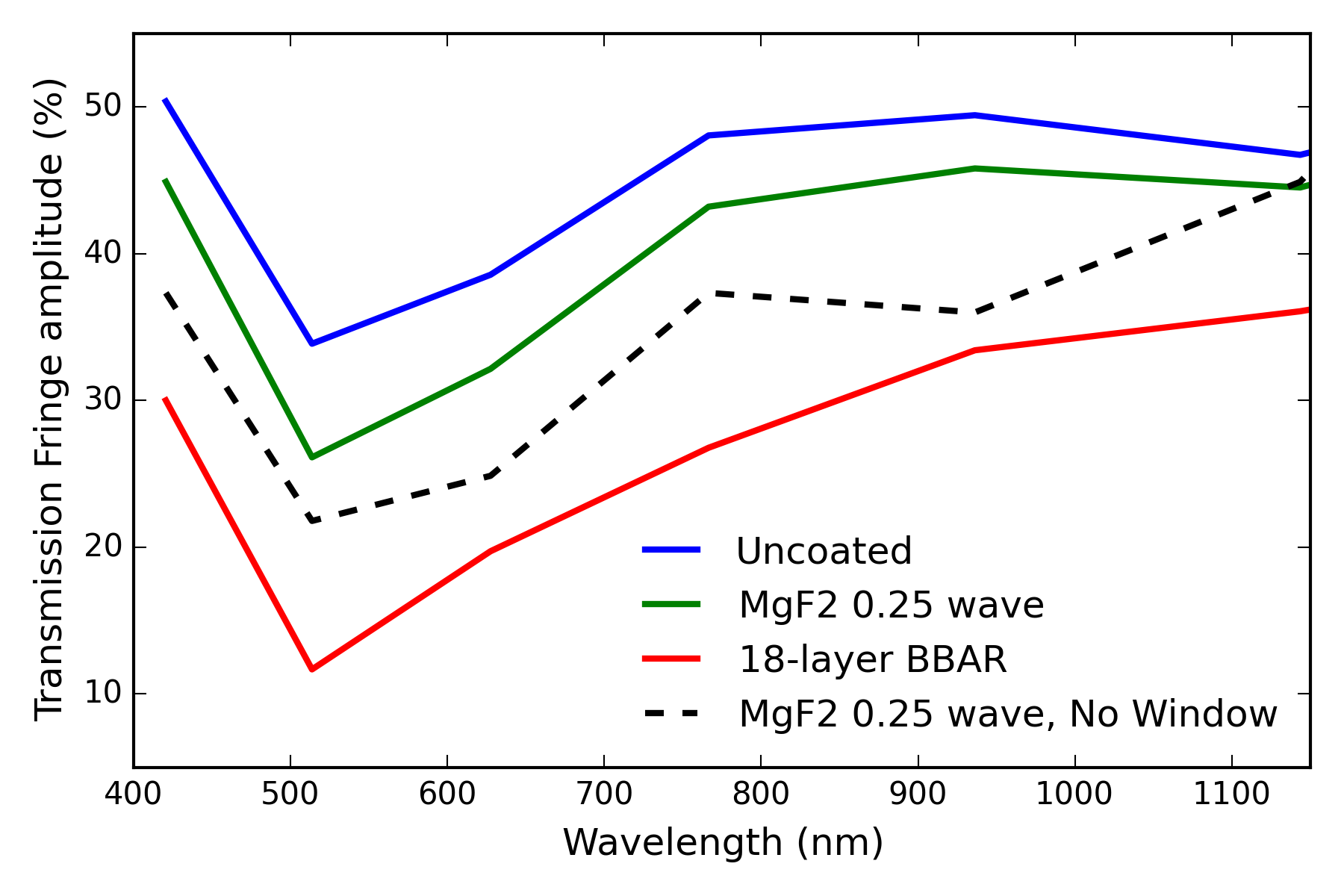}
}
\end{tabular}
\caption
{ \label{fig:visp_pcm_window_coating_fringe_impact_transmission} 
The polarization fringe amplitudes for the optic transmittance using various scenarios for our visible wavelength optimized quartz-based retarder. The Berreman predicted Mueller matrix was computed at spectral sampling of $\lambda$ / $\delta \lambda$ of 2,000,000 and subtracted from the ideal perfect retarder design. Fringe amplitudes were computed in wavelength bins of roughly 100nm. Blue shows no coating on the window exterior interfaces.  Green shows a single layer isotropic MgF$_2$ coating.  Dashed black shows the same coating but with both Infrasil windows removed.   Red shows an 18-layer BBAR coating from Infinite Optics on the window exterior interfaces.  See text for details.}
\vspace{-3mm}
\end{wrapfigure}

The amplitude and spectral frequency of the fringes depends strongly on the refractive index mis-match between air, the windows and the two indices in the crystals. Anti-reflection coatings can be useful to reduce fringe amplitudes, but they must be considered carefully for other risks.  The index mis-match between the CaF$_2$ and the crystal MgF$_2$ e- and o- beams is roughly (0.045, 0.055) at 854nm wavelength. 

Two of the DKIST calibration retarders are made of crystal quartz with Heraeus Infrasil 302 fused silica cover windows \cite{Sueoka:2014cm,2014SPIE.9147E..0FE,Sueoka:2016vo}. For the fused silica to crystal quartz e- and o- beam interface this refractive index mismatch is (0.085, 0.107) at 854nm wavelength. If the quartz crystals were left uncoated, the fringes in these retarders would be substantially worse. The quartz crystals are anti-reflection coated with isotropic MgF$_2$, substantially reducing the internal fringe amplitude around the anti-reflection coating design wavelength.

Anti-reflection coatings come with benefits as well as risks.  Each crystal must be polished to a specific retardance in subtraction. If a coating run fails, as some did for DKIST, the damaged crystals must be re-polished. For such thin crystals, this can result in loss of the crystal at significant schedule and cost impact. Coatings also absorb heat and add complication for the DKIST calibration retarders in the 300W beam. We can use the Berreman calculus to predict fringe behavior for a variety of anti-reflection coatings. In Figure \ref{fig:visp_pcm_window_coating_fringe_impact_transmission} we show the fringe amplitude for the [0,0] Mueller matrix element ($II$ , transmission) for the visible-wavelength optimized quartz achromatic calibration retarder.  

In all models, there is 75.5nm physical thickness of {\it isotropic} MgF$_2$ as an anti-reflection coating on internal crystal interfaces. This corresponds to a quarter-wave of optical path at 525nm central wavelength, the common single-layer AR coating.  The oil thickness is set to 10$\mu$m between all six crystals and both window internal interfaces.

To easily show the fringe amplitudes, we computed the Berreman models at spectral sampling of $\lambda$/$\delta \lambda$ of 2,000,000.  These models were then subtracted from the theoretical Mueller matrix to compute residual fringe amplitudes. These residual fringe amplitudes were then binned spectrally by a factor of 400,000 to create smooth, low resolution wavelength dependence that represents the envelope for polarization fringes. Fringe frequencies have periods $<<$ nm but the amplitudes have envelopes that effectively follow the anti-reflection coating behavior as seen in Figure \ref{fig:visp_pcm_window_coating_fringe_impact_transmission}. 

The blue curve shows fringe amplitudes for a Berreman model with an uncoated Infrasil 302 fused silica windows.  The green curve shows a quarter-wave {\it isotropic} MgF$_2$ anti-reflection coating on the Infrasil 302 window external surfaces (as well as internal) with a central coating wavelength of 525nm (75.5nm physical thickness). The red curve shows fringe amplitudes when an 18-layer broad-band anti-reflection (BBAR) formula is coated on the exterior cover window surfaces interfaced with air. The BBAR coating was $\sim$1\% reflectivity at all wavelengths in the 380nm to 1100nm wavelength range.  

The black curve shows the fringe amplitude for a model without any Infrasil cover windows at all. In this no cover window Berreman model, there are single-layer AR coatings on all quartz crystal internal interfaces as seen in Figure \ref{fig:coatings_for_DKIST_retarder}. The exterior-facing quartz crystals have the nominal 525nm central wavelength isotropic MgF$_2$ anti-reflection coating. The internal interfaces have this same coating as well as the index matching oil. Without cover windows, the fringe period would be roughly 3 times larger as the optic is roughly 3 times thinner optically.  The amplitude in Figure \ref{fig:visp_pcm_window_coating_fringe_impact_transmission} is higher than the red curve becuase the 18-layer BBAR coating is a much better anti-reflection coating than a single layer of isotropic MgF$_2$.

\subsection{Anti-reflection Coatings Impact Fringes}

As an example of the usefulness of this tool, we computed several fringe models using various types of anti-reflection coatings on the various interfaces. Fringe amplitudes are directly impacted by the electric field amplitude in the backward traveling waves.  In many-crystal stack retarders, the anti-reflection coating properties become very critical in achieving low fringe amplitudes.

\begin{wrapfigure}{r}{0.65\textwidth}
\centering
\vspace{-6mm}
\begin{tabular}{c} 
\hbox{
\hspace{-1.2em}
\includegraphics[height=7.5cm, angle=0]{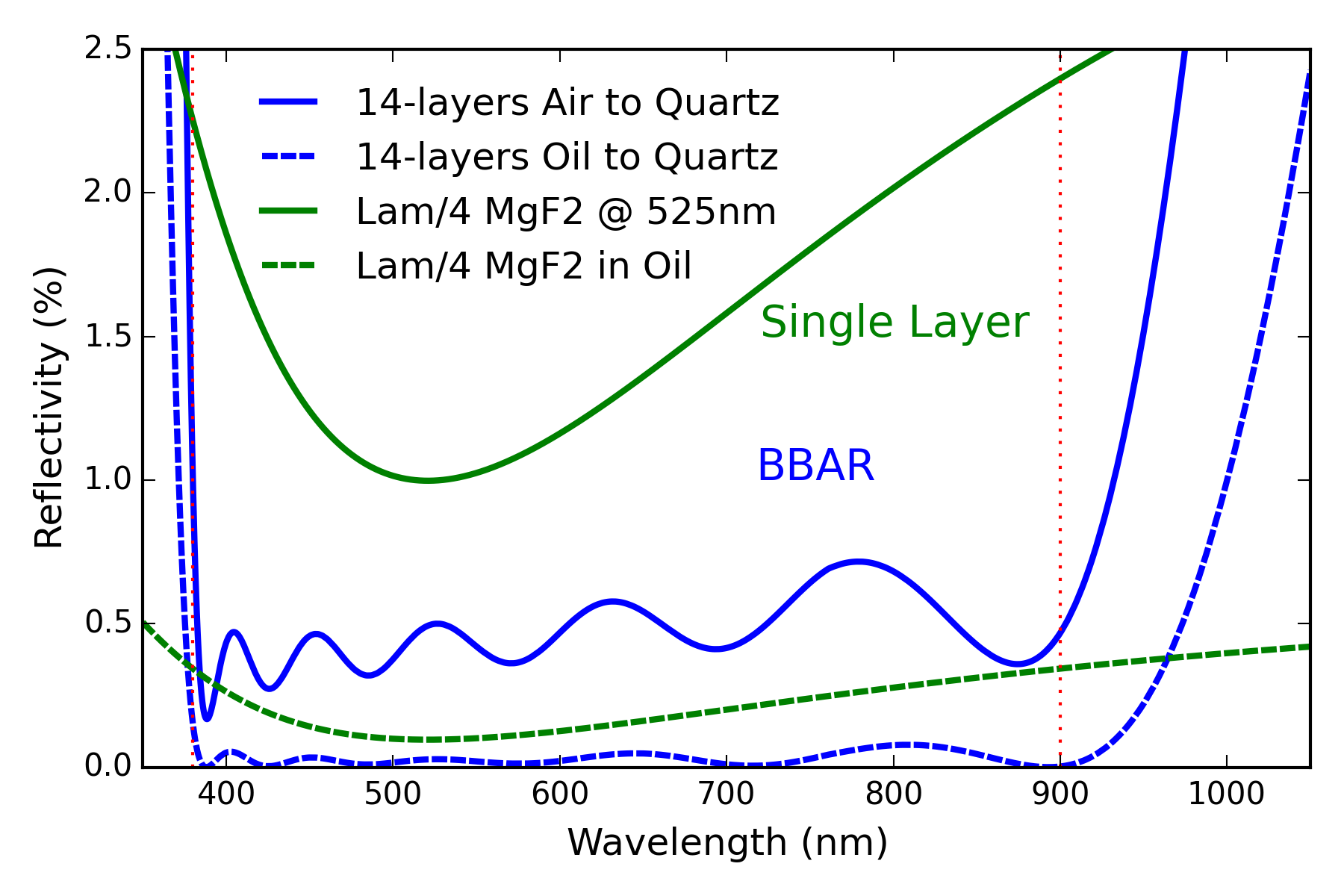}
}
\end{tabular}
\caption[Coatings for DKIST Retarders] 
{ \label{fig:coatings_for_DKIST_retarder} 
The reflectivity for coatings used in the Berreman fringe calculations. Green curves show a simple single-layer of isotropic MgF$_2$ at quarter-wave thickness for a central wavelength of 525nm.  The dashed green line shows this coating interfaced to the oil at n$\sim$1.3.  The solid blue curve shows a broad-band AR coating with 14 layers optimized for the 380nm to 900nm bandpass on quartz crystal interfaced to air.  The dashed blue curve shows 14 layers of the same material but with slightly different layer thicknesses optimized for the crystal quartz to index matching oil interface.  See text for details. }
\vspace{-1mm}
 \end{wrapfigure}

Figure \ref{fig:coatings_for_DKIST_retarder} shows example coatings used in our fringe calculations. For the visible wavelength quartz-based retarders used in the 380nm to 900nm wavelength range, we applied a traditional anti-reflection coating as a simple quarter-wave thickness of (isotropic) MgF$_2$ with a central wavelength of 525nm. This corresponds to  a physical thickness of 94nm and is shown as the solid green line for the air to quartz crystal interface in Figure \ref{fig:coatings_for_DKIST_retarder}. As we have a refractive index matching oil in between the interior crystal interfaces, we also show this same single-layer MgF$_2$ coating as the dashed-green line in Figure \ref{fig:coatings_for_DKIST_retarder} when the interface is oil to quartz crystal.  This simple quarter-wave MgF$_2$ coating gives a minimum reflectivity at the central 525nm wavelength of about 1\% for the air interface and 0.1\% for the oil interface.  For reference, the crystal quartz to oil interface has a reflectivity of about 0.9\% at 380nm falling to 0.7\% at 1500nm wavelength without coatings.  Uncoated crystal quartz has a $\sim$5\% reflection to air at 380nm wavelength falling to about 4.5\% reflection at 1500nm wavelength.

Broad-band anti-reflection (BBAR) coatings offer substantial reduction of fringe amplitude over a wide wavelength range.  This translates to improved optical performance when these retarders are used for modulation or calibration with high spectral resolution astronomical instruments. As an example, we show in Figure \ref{fig:coatings_for_DKIST_retarder} a 14-layer coating formula that has been optimized for the air-quartz interface in the 380nm to 900nm wavelength range as the solid black line.  Another 14-layer BBAR coating formula using the same materials has been optimized for the oil-quartz interface as the dashed blue line with slightly different layer thicknesses.  As expected, these BBAR coatings reduce the reflectivity substantially.  The improvement compared to the simple single-layer isotropic MgF$_2$ coating is a factor of 2x at the central wavelength of 525nm but is up to 5x improvement across the specified bandpass range. Wide wavelength ranges are typically required of astronomical spectropolarimeters as instruments usually cover the visible and near-infrared bandpasses.

\begin{figure}[htbp]
\begin{center}
\vspace{-2mm}
\hbox{
\hspace{-1.0em}
\includegraphics[height=11.2cm, angle=0]{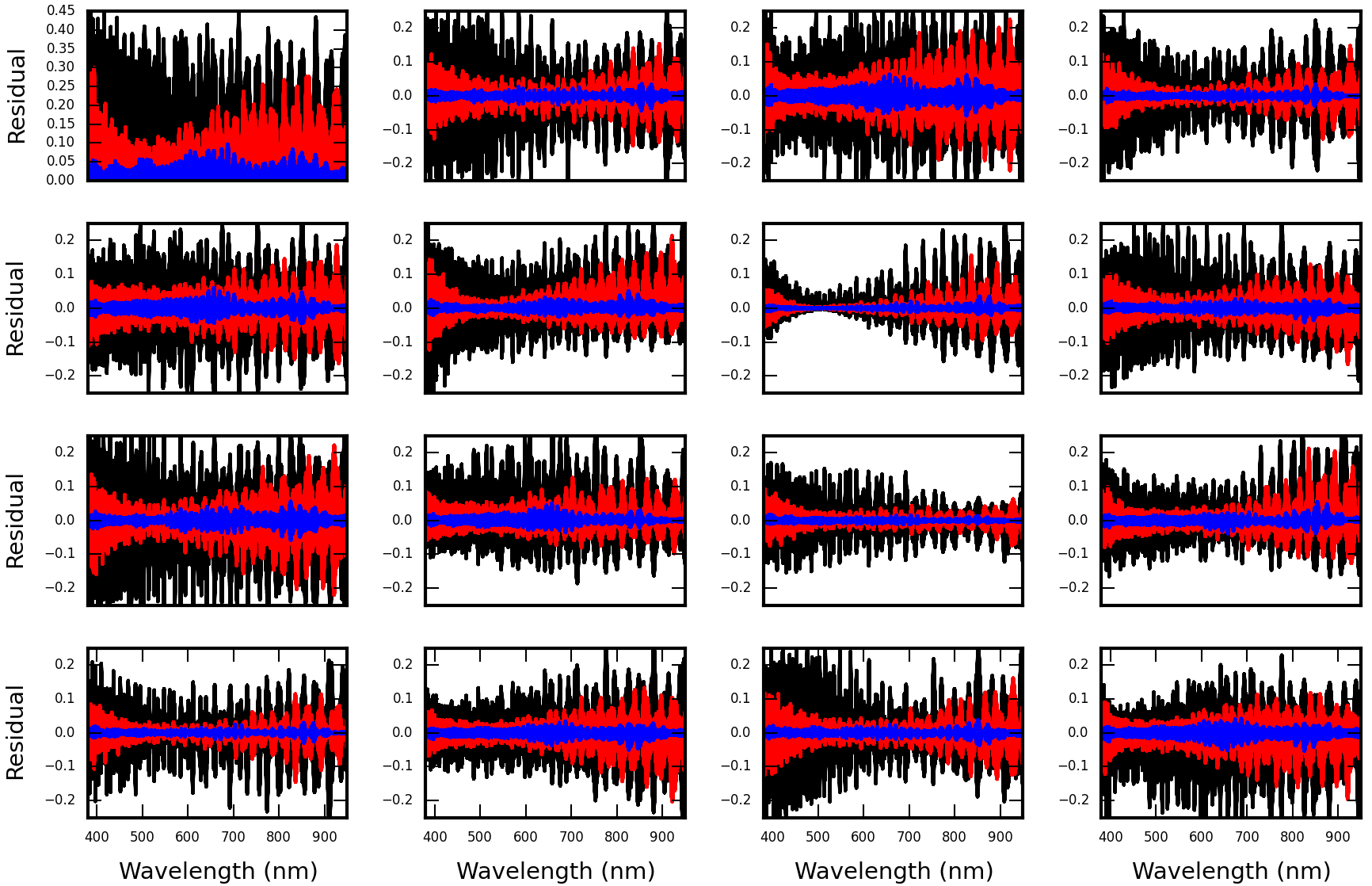}
}
\caption{The difference between the ideal Mueller matrix (no fringes) and the Berreman computed Mueller matrix (with fringes). The ideal Mueller matrix is computed as a stack of ideal linear retarders following Table \ref{table:ViSP_SAR_Design}. This ideal Mueller matrix is subtracted from the Berreman-calculated Mueller matrix, leaving only residual fringes. This calculation does not use the Infrasil fused silica windows nor the coatings and oil layers required for such a window listed in Table \ref{table:ViSP_SAR_Design}. Black shows fringes when a single-layer MgF$_2$ coating is used only on the exterior air-quartz interface and no coatings are applied at the oil-quartz interfaces. Red shows single-layer MgF$_2$ coatings on all surfaces, oil and air interfaces. The blue curve shows the 14-layer BBAR coatings from Figure \ref{fig:coatings_for_DKIST_retarder} applied to all interfaces.  }
\label{fig:ViSP_PCM_Fringe_Residual}
\vspace{-8mm}
\end{center}
\end{figure}

These coatings cause a strong decrease in polarization fringe amplitudes and they are easily included in the Berreman formalism. Figure \ref{fig:ViSP_PCM_Fringe_Residual} shows the difference between the theoretical Mueller matrix and the Berreman-computed Mueller matrix including fringes.  The theoretical Mueller matrix is computed using ideal linear retarder Mueller matrices following the crystal orientations in Table \ref{table:ViSP_SAR_Design}.  There are no fringes in this calculation.  We then compute Mueller matrix models with the Berreman formalism for the same stack of crystals but now we include various AR coating layers and refractive index matching oil layers as in Table \ref{table:ViSP_SAR_Design}.  The Berreman formalism computes a Mueller matrix matching theory, but now including all polarization and intensity fringes. 

The black curve in Figure \ref{fig:ViSP_PCM_Fringe_Residual} shows fringes when a single-layer MgF$_2$ coating is used only on the exterior air-quartz interface and no coatings are applied at the oil-quartz interfaces. Transmission fringes are up to 40\% with the $I$ to $QUV$ elements up to 20\% and retardance elements at roughly 0.2.  The red curve shows the nominal design with single-layer MgF$_2$ coatings on all interfaces, oil and air.  These additional internal coatings reduce the fringes most significantly near the 525nm central wavelength. The fringe reduction between black and red curves shows that coating the internal interfaces is critical even though we do have an oil at refractive index 1.3. We note that a refractive index 1.5 oil would have been preferable.  However, searches for oils that could meet all requirements including 30+ year lifetime without service, no degradation with UV flux exposure, no absorption bands from 380nm to 5000nm wavelength and additional constraints were unsuccessful. 

We then simulate fringes amplitudes with BBAR coatings on all interfaces as the blue curves of Figure \ref{fig:ViSP_PCM_Fringe_Residual}.  The BBAR coatings do produce a significant improvement at all wavelengths but the improvement is strongest at wavelengths away from the 525nm central wavelength of the single-layer coating. The BBAR coatings further reduce the fringes by another factor of several, showing strong benefit.

\subsection{Expected Fringe Periods for AO-assisted DKIST Instruments}

We use the Berreman framework to predict the expected fringe periods for the major transmissive optics in the DKIST optical path.

\begin{figure}[htbp]
\begin{center}
\vspace{-3mm}
\hbox{
\hspace{-1.5em}
\includegraphics[height=11.8cm, angle=0]{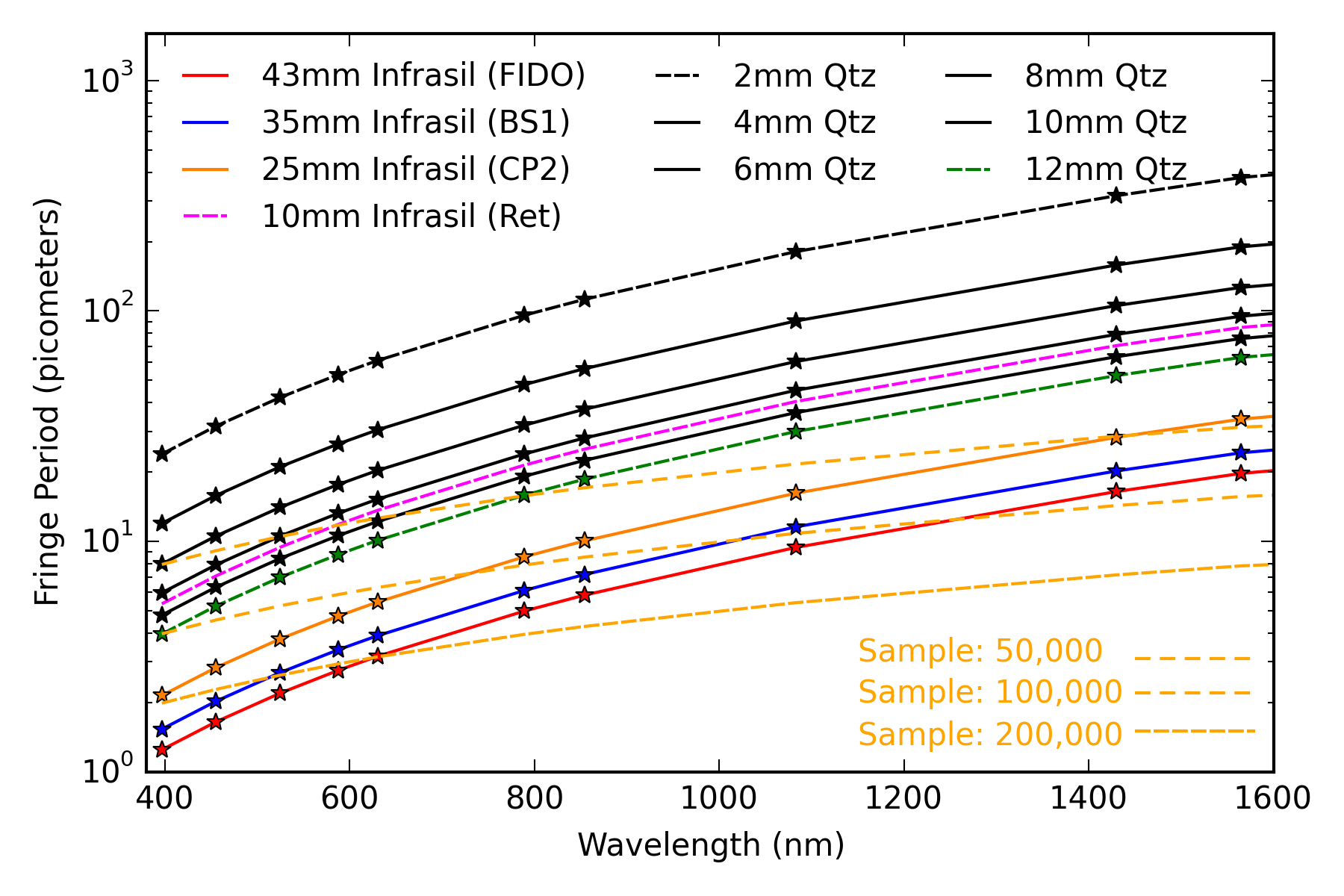}
}
\caption[Fringe Frequency for DKIST] 
{ \label{fig:fringes_for_dkist} 
The predicted fringe periods created by the DKIST transmissive optics for instruments using the AO-assisted beam. The dashed orange curves show spectral sampling at typical DKIST rates ($\lambda$/$\delta \lambda$) of 50,000 to 200,000. A typical calibration observation would use all listed optics including one to four dichroic BSs, the WFS-BS1, six quartz crystals and a 25mm thick window in the calibration polarizer (CP2). Beam f/ numbers range from f/13 to collimated so fringe amplitude predictions are not listed. See text for details.}
\vspace{-6mm}
\end{center}
\end{figure}

In the coud\'{e} lab, the AO-corrected beam is then distributed by the dichroic beamsplitters of the Facility Instrument Distribution Optics (FIDO) for simultaneous use of several instruments.  These FIDO dichroics are re-configurable and will have special coatings. Most instruments observe transmitted through one to several beam splitters.  However, the optics and coatings have yet to be procured as of mid 2017 so exact prediction of fringe amplitude and wavelength range will be performed later. 

There are two 10mm thick Infrasil 302 windows originally included in the design of the DKIST retarders \cite{Sueoka:2014cm, 2014SPIE.9147E..0FE, Sueoka:2016vo}. During polarization calibration of an instrument, there would be four of these 10mm thick windows in the beam. There are two retarders, one at Gregorian focus for calibration and one acting as a modulator each with two windows. We have recently removed the windows from all retarders to mitigate fringes following successful testing for other performance parameters (transmitted wavefront error, beam deflection, uniformity).  Each of the DKIST retarders consists of six crystals of roughly 2mm thickness each and there will be a fringe at all thickness combinations.  During polarization calibration, there is also a window included in one of the wire grid polarizer assemblies we call Calibration Polarizer 2 (CP2). This window is presently designed as 25 mm thickness of Heraeus Infrasil 302. This window will absorb long wavelengths to provide thermal mitigation for the retarders. 

We compile the predicted fringe period as a function of wavelength in Figure \ref{fig:fringes_for_dkist}. Fringes for BS1, the FIDO dichroics, the CP2 window, retarder crystals and the now-removed retarder windows (Ret) all produce resolvable fringes. Symbols in Figure \ref{fig:fringes_for_dkist} mark several common observing wavelengths. The thicker the optic, the smaller the fringe period.  With the thicker optics, the required spectral resolution required to resolve the fringe also increases. As an example, at 396.8nm wavelength, the Infrasil substrates for the FIDO beam splitters produce a fringe with a period of 1.25pm. The required spectral sampling to measure this fringe at two points per wave would be $R = 638,000 = \frac{\lambda}{\delta\lambda}$.  For the same 43mm thick FIDO substrate measured at a wavelength of 1565 nm the fringe period is 19.7 nm and two-point sampling would require spectral sampling at $R=159,000$.  The retarder crystals produce fringes that are easily resolved by all DKIST high spectral resolution instruments.  The various windows also produce fringes that are marginally to fully resolved.

\section{Summary}

We presented polarization fringe models for several thick-crystal stack retarders using the Berreman calculus framework. This framework is able to compute fringes for an arbitrary number of layers of arbitrarily oriented birefringent materials and varying incident angles while computing all intensity and polarization artifacts. The theory matches simple analytic formulas exactly in the limit of no coherent interference. We applied the Berreman framework to isotropic windows and coating materials to verify consistent outputs between our software and common optical modeling software tools such as Zemax and TFCalc. Our results agree with the analytic formalisms common in the thin film industry. We also reproduced results from several papers in the astronomical literature. Predictions for sapphire and quartz bi-crystalline achromats with and without air gaps match observations of fringe spectral periods for a retarder in use at the Dunn Solar Telescope on a high resolution spectropolarimeter, SPINOR. Predictions also match laboratory experiments performed on simple windows and a quartz retarder. 

The DKIST polarimetric optics and AO-assisted beamsplitter windows were presented and analyzed. With the Berreman calculus, we showed how anti-reflection coatings, refractive index matching oil and cover windows impact fringe predictions. We used this tool to show how DKIST retarder design decisions can be informed by fringe analysis. This tool gives quantitative predictions for fringes as changes to the optic thickness, coatings, temperature stabilization and many other factors are considered. This new tool is very useful for providing theoretical understanding of fringe origin and dependencies in many-crystal retarder designs. Having this tool for predicting polarization fringe spectral periods and amplitudes in relatively slow beams brings quantitative analysis to the design and trade studies for large crystal retarders in use at astronomical telescopes. It also provides simple scaling relations when considering how fringes depend on thermal issues, heat deposited by absorptive coatings, added optical interfaces and other potential couplings between optical and various thermal, mechanical and environmental considerations.

\section{Acknowledgements}

This work was supported by the DKIST project. The DKIST is managed by the National Solar Observatory (NSO), which is operated by the Association of Universities for Research in Astronomy, Inc. (AURA) under a cooperative agreement with the National Science Foundation (NSF). We thank Dr. David Elmore for his assistance, guidance and insight into the long history of work on the DKIST project.  We thank Dr. Christian Beck for his expert observation and data analysis at the Dunn Solar Telescope with SPINOR.  This research made use of Astropy, a community-developed core Python package for Astronomy (Astropy Collaboration, 2013).  http://www.astropy.org.


\bibliography{ms_ver03} 			
\bibliographystyle{spiebib}		

\end{document}